\documentclass[12pt]{spieman}  
\usepackage{amsmath,amsfonts,amssymb}
\usepackage{graphicx}
\usepackage{setspace}
\usepackage{tocloft}
\usepackage{lineno}
\usepackage{threeparttable}
\usepackage{aas_macros, rotating}
\providecommand{\sorthelp}[1]{}

\title{Simulation of the Far-Infrared Polarimetry Approach Envisioned for the PRIMA Mission}
\author[a,*]{C. Darren Dowell}
\author[a]{Brandon S. Hensley}
\author[b]{Marc Sauvage}

\affil[a]{Jet Propulsion Laboratory, California Institute of Technology, 4800 Oak Grove Drive, Pasadena, CA 91109, USA}
\affil[b]{Universit\'e Paris-Saclay, Universit\'e Paris Cit\'e, CEA, CNRS, AIM, 91191, Gif-sur-Yvette, France}

\cftpagenumbersoff{figure}
\cftpagenumbersoff{table} 
\begin{document} 
\maketitle

\begin{abstract}
Interest in the study of magnetic fields and the properties of interstellar dust, explored through increasingly capable far-IR/submillimeter polarimetry, along with maturing detector technology, have set the stage for a transformative leap in polarization mapping capability using a cryogenic space telescope.  We describe the approach pursued by the proposed PRobe far-Infrared Mission for Astrophysics (PRIMA) to make ultra-deep maps of intensity and polarization in four bands in the 92—235\,$\mu$m\ range.  A simple, polarimetry-optimized PRIMAger Polarimetric Imager (PPI) is designed for this purpose, consisting of arrays of single-polarization Kinetic Inductance Detectors oriented with three angles which allow measurement of Stokes $I$, $Q$, and $U$ in single scans.  In this study, we develop an end-to-end observation simulator to perform a realistic test of the approach for the case of mapping a nearby galaxy.  The observations take advantage of a beam-steering mirror to perform efficient, two-dimensional, crossing scans.  Map making is based on ‘destriping’ approaches demonstrated for Herschel\footnote{Herschel is an ESA space observatory with science instruments provided by European-led Principal Investigator consortia and with important participation from NASA.}/SPIRE and Planck.  Taking pessimistic assumptions for detector sensitivity including $1/f$ noise, we find excellent recovery of simulated input astrophysical maps, with $I$, $Q$, and $U$ detected at near fundamental limits.  We describe how PPI performs detector relative calibration and mitigates the key systematic effects to accomplish PRIMA polarization science goals.
\end{abstract}


{\noindent \footnotesize\textbf{*}C. Darren Dowell,  \linkable{charles.d.dowell@jpl.nasa.gov}}

\begin{spacing}{1}     

\section{Introduction}

The advent of sensitive observations of the polarized far-infrared (FIR) to submillimeter emission from interstellar dust over the last two decades has had broad impact across astrophysics. Dust grains preferentially align with their short axes parallel to the local magnetic field, and so the dust polarization angle is a measure of the magnetic field orientation projected onto the plane of the sky \cite{Purcell:1975,Andersson2015}. The relationship between dust, gas, and magnetic fields revealed through such observations has informed models of star formation (e.g., \citenum{planck2015-XXXV, Pattle2019, Pillai2020, Arzoumanian2021, Ward-Thompson:2023}), the structure of magnetic fields (e.g., \citenum{planck2016-XLII, Fissel:2019, Borlaff2023, Surgent:2023}), interstellar turbulence (e.g., \citenum{Soler2016, Caldwell:2017, planck2016-l11B}), interstellar filaments (e.g., \citenum{Clark:2015, planck2014-XXXIII, Huffenberger:2020, Ching:2022, Ngoc:2023}), dust composition (e.g., \citenum{Guillet:2018, Draine:2021, Hensley:2023, Ysard:2024}), and grain alignment (e.g., \citenum{Hildebrand:1999, Tram2021b, Hoang2023, LeGouellec:2023}) across environments from star-forming cores to the diffuse interstellar medium (ISM) of the Milky Way to resolved nearby galaxies, with some recent observations even at cosmological redshifts \cite{Geach:2023}.

Neglecting circular polarization, which is rare in this astrophysical context, the polarization information consists of three elements to be measured at each of many points on the sky: either a total intensity, a polarization fraction, and a polarization angle; or alternatively the first three components of the Stokes vector $(I, Q, U)$ where $I$ is the total intensity and $Q$ and $U$ specify the linearly-polarized intensity.  A variety of strategies have emerged for mapping those three quantities for FIR continuum radiation, involving choices of detectors, optics, and motion of the optical boresight on the sky. Measurement of the polarization of light is inherently differential, so all methods involve subtraction of detector signals, either simultaneous within common optical paths or separated in time.

There exist three broad families of detectors: (1) Dual-polarization detectors. These use two co-located elements sensing orthogonal linear polarizations, e.g., Planck \cite{Lamarre2010}, POLARBEAR \cite{Barron2014}, or the SPICA/B-BOP design \cite{Adami2019}. Dual-polarization detectors directly measure $I$ and one polarization Stokes parameter, and they mitigate certain systematic effects such as common-mode drifts and pointing error.  (2) Single-polarization detectors. Though they are simpler to design, there are relatively few instruments employing such detectors, e.g., BLASTPol, \cite{Fissel2013,Galitzki2014} and the SKIP concept \cite{Johnson2014}.  (3) Total power detectors. These detectors -- such as in SCUBA+POL-2 \cite{Friberg2016}, HAWC+ \cite{Harper2018}, or the NIKA suite of instruments \cite{Ritacco2017} -- are made capable of measuring polarization using a polarizing filter in the upstream optics. This assembly is functionally equivalent to single-polarization detectors.
As a sub-category of (3), several instruments use the reflected as well as transmitted radiation from the polarizing filter, and two detectors, to measure $I$ and one Stokes polarization parameter, similar to dual-polarization detectors.  However, this approach takes up more space at the focal plane and necessitates good optical alignment to avoid effects such as differential focus.
Note that none of these approaches is capable of measuring, in a single instant in time, the full Stokes vector for one point on the sky.

To recover all of $I$, $Q$, and $U$, two strategies have been employed. The first is to fill the focal plane with pixels having their polarization-selective directions oriented with enough angles to fully specify the Stokes vector
so that all of the information can be obtained at almost the same time, but with different pixels, by shifting the optical boresight on the sky. The second is use of a polarization rotator, i.e., a half-wave plate (HWP), to effectively rotate the incoming polarization so that all of the information can be obtained with the same pixels, but not at the same time.
A HWP mitigates effects such as gain errors and provides opportunity for additional signal modulation (and of the component of primary interest:  the polarized one).
This explains why almost all of the polarimetric instruments implemented so far, except on Planck, have introduced a rotating HWP in their optical path.  Use of a HWP often comes with a need to accurately characterize ``instrumental polarization.''

The PRIMAger instrument\cite{Ciesla25}, designed for the PRobe far-Infrared Mission for Astrophysics (PRIMA\cite{Moullet:2023,Meixner:2024}) takes a simple approach, using single-polarization detectors oriented at three angles and no polarization modulating device. In this study, we demonstrate through simulations that the instrument and observatory designs allow total intensity and polarization to be measured with excellent sensitivity as a result of scanned observations.  The Kinetic Inductance Detectors (KIDs\cite{Baselmans2022}) in PRIMAger, operating at the background limit for a cryogenic telescope, enable the driving polarimetric science case of PRIMA -- deep polarimetric mapping of nearby galaxies -- even in the presence of $1/f$ noise in the detectors at a pessimistically high level.

This paper is organized as follows. In Section~\ref{sec-primapol}, we describe the design with which PRIMA intends to measure polarization and the fundamental measurement requirements.  Section~\ref{sec-mapping} details the algorithm for processing PRIMA detector samples to produce maps of intensity and polarization.  Section~\ref{sec-sim} describes an end-to-end simulation of example PRIMA polarization observations and analyzes the output products.  Sections~\ref{sec-gain} and \ref{sec-beam} address certain optical design requirements and calibrations in support of PRIMA polarimetry, and finally Section~\ref{sec-conclusion} summarizes the key results of this study.

\section{Far-IR Continuum Polarimetry with PRIMA\label{sec-primapol}}

PRIMA and PRIMAger are designed to enable polarimetric imaging in at least two FIR bands with a polarization sensitivity of $\leq$0.03\,MJy/sr ($5\sigma$) in each band, per $27^{\prime\prime} \times 27^{\prime\prime}$ (FWHM) resolution element and over a 10\,sq.\,arcmin region, within a 2\,hr observation.

To achieve this science requirement, PRIMA will feature a 1.8\,m diameter, cryogenic telescope delivering a field of view of $42'\times24'$ to its multi-wavelength, broad-band ($R = \lambda/\Delta\lambda = 4$--$10$) mapping instrument PRIMAger. The PRIMAger Polarimetric Imager (PPI) half of the instrument offers four focal plane detector arrays operating simultaneously with $R \approx 4$ centered at $\lambda$ = 92, 125, 165, and 235\,$\mu$m.  Each detector array has an instantaneous field of view of $\sim\! 4'\times 4'$. Two mapping modes are foreseen for PRIMAger: for objects smaller than the telescope field of view, a beam-steering mirror (BSM) moves the instrument field of view within the telescope field of view, while for degree-sized fields, a combination of scanning with the observatory and the BSM will be employed.

The BSM design is based on a similar device operating on Herschel/PACS \cite{Krause2006}. It is a two-axis mirror allowing full exploration of the telescope field of view in any direction. The current design of the BSM allows for a maximum speed of 1000\,$^{\prime\prime}$/s on the sky and maximum acceleration of 10,000\,$^{\prime\prime}$/s$^2$.
The BSM also hosts a location for an internal calibration source that is used to track detector gain variations during operations and to calibrate detector nonlinearity, as was done for Herschel/SPIRE \cite{Bendo2013}.

The focal plane arrays consist of KID detector pixels, where each pixel is the combination of a KID and a microlens that focuses the incoming light on the collecting or absorbing element of the detector. Pixels are arranged in a hexagonal pattern, achieving maximum packing, and the optical design is such that an $F\lambda$ sampling is realized. A single pixel is sensitive to a single polarization direction; therefore to sample the polarization information completely, pixels sensing three different orientations, each rotated from the other by 120$^{\circ}$, are placed in the focal plane, in a mixed pattern.

The PRIMA approach to polarimetry is most similar to BLASTPol \cite{Fissel2013, Galitzki2014}, with two differences:  1) PPI uses three polarization angles in each detector array, instead of two, to allow measurement of $(I, Q ,U)$ in single scans; 2) PPI omits the HWP used in BLASTPol.  Located near Sun-Earth Lagrange Point 2, PRIMA should have excellent thermal stability and therefore sufficient stability in detector gain to permit effective differencing over hour-long observations.  The temporal stability of background power — dominated by astrophysical sources — as well as the scan modulation provided by the BSM eliminate the need for additional HWP modulation.  An internal calibration source is used periodically to maintain this gain stability for the duration of the mission.

The capabilities of PPI are well-suited to making polarimetric maps of nearby galaxies. Total intensity surveys undertaken by Herschel, e.g., the Key Insights on Nearby Galaxies: A Far-Infrared Survey with Herschel (KINGFISH\cite{Kennicutt:2011}) and the Herschel Reference Survey (HRS\cite{Boselli:2010}), demonstrate the availability of large ($\gtrsim 20$ arcmin$^2$) nearby galaxies spanning a range of metallicities and star formation rates that have bright FIR dust emission. In the KINGFISH sample, emission from galaxies was able to be recovered down to a typical surface brightness limit of 1.6\,$L_\odot$\,pc$^{-2}$ (Ref.~\citenum{Aniano:2020}), or roughly 1\,MJy/sr at 250\,$\mu$m. As resolved observations of nearby galaxies with SOFIA/HAWC+ reveal polarization fractions of order a few percent \cite{Lopez-Rodriguez:2022}, this translates to roughly 0.03\,MJy/sr of polarized intensity in the faintest pixels (though we note that faint regions often have higher polarization fractions; see, e.g., \citenum{Jones:1992, planck2014-XIX, Lopez-Rodriguez:2022}). Dust models predict ratios in the dust polarization fraction between 235 and 92\,$\mu$m ranging from nearly one to factors of two or more \cite{Draine:2009, Guillet:2018, Hensley:2023, Ysard:2024}. High-signal-to-noise measurements of the polarization fraction ($\sigma(p) \lesssim 0.5\%$) at multiple frequencies promise strong discrimination among them.

\section{Polarization Map Reconstruction} \label{sec-mapping}

\subsection{Overview} \label{sec-map-intro}

The functional requirement of PPI is to make spatially resolved maps of Stokes $I$, $Q$, and $U$ (total and linearly-polarized intensity) in multiple FIR bands.  Like past far-IR/(sub)mm missions that use scanning to map the sky (e.g., \citenum{Griffin2010,Poglitsch2010,Tauber2010}),
PPI senses intensity by recording the change in power absorbed in each detector as its far-field beam is swept across the observational target.  For polarization-sensitive detectors such as in PPI, the three Stokes parameters $I$, $Q$, and $U$ contribute to the observed signal, so a single scan with a single detector is not sufficient to uniquely distinguish the terms.  At a minimum, three scans with detector(s) oriented at three well-separated polarization angles are required.  In practice, PPI covers each pixel in the target mapping region with highly redundant measurements; in that region, $I$, $Q$, and $U$ are well determined, and many measurements are used to average down photon and detector noise.  Below, we describe a least-squares approach to combining the measurements to generate spatial maps of $I$, $Q$, and $U$.

PPI KIDs do not have a clear reference to the (hypothetical) case of zero input radiation power, similar to the situation for bolometers (e.g., \citenum{SPIRE2016}).  The KID signal resides on an electronic ``baseline'' which is not of astrophysical interest but must be removed to generate sensitive maps.  The detector baselines are independent and large, so an important role of the map maker is to ``destripe'' the maps by adjusting the subtracted baselines to minimize the variance of the redundant measurements.  Destriping is most effective when scans are ``crosslinked,'' utilizing multiple directions crossing at $\gtrsim 30^\circ$ angles.  Due to the nonzero baselines for zero input power, PPI measures only {\em relative} $I$, $Q$, and $U$ over the mapped area, and a post-processing step is needed to set the absolute additive flux density reference, typically using the edge of the mapped area.

\subsection{Signal Model}

Consider a position $p$ on the sky at which there is an astrophysical signal characterized by the Stokes vector ($I_p$, $Q_p$, $U_p$). The response $S$ of a detector with gain $g$, polarization efficiency $\epsilon$, and polarization orientation $\theta$ to this signal is (see, e.g., \citenum{Kurki-Suonio:2009})

\begin{align}
    S_{j,k} &= g\{I_p + \epsilon[Q_p\cos2\left(\theta-\phi_j\right) + \nonumber \\ &U_p\sin2\left(\theta-\phi_j\right)]\} + b_j + n_{j,k} \label{eq-sigmodel}
\end{align}
where the indices $j$ and $k$ indicate the scan number and sample number, respectively; $\phi_j$ is the line-of-sight rotation angle of the instrument, which we assume is constant over a single scan; $b_j$ is the electronic baseline of the detector signal; and $n_{j,k}$ is the noise timestream.

The detector gain is assumed to be linear and constant in time either as a natural behavior of the detector, or, more likely, as a result of separate gain characterization, tracking, and correction.  The detector polarization response, characterized by $\epsilon$ and $\theta$, should be intrinsically time stable and is assumed to be known from ground testing. We have assumed for now that $I$, $Q$, and $U$ represent the convolution of the astrophysical signal with the instrument angular response and are solely dependent on position $p$. In Section~\ref{sec-beam}, we consider detector-dependent variations in $I_p$ which arise from variations in instrument optics.

PRIMA and PRIMAger have off-axis optical elements likely to induce instrumental polarization at the few percent level.  Since there is no internal polarization modulator, this instrumental polarization is not visible directly in the signal timestreams.  To first order, the induced polarization combines with $g$, $\epsilon$ , and $\theta$ and slightly modifies their values.  Second-order effects of instrumental polarization and other systematics will be the subject of a future detailed design study.

The noise $n$ has an expectation value of zero. We define the noise weight $1/\sigma^2$ based on $n$. For white noise, $\sigma^2_{j} \equiv \langle n_{j}^2 \rangle$.

\subsection{Least-squares solution for (I,Q,U)\label{sec-solution}}

The map maker demonstrated here operates by iteratively performing a least-squares solution first for the $I$, $Q$, and $U$ maps given assumed detector baselines, and then the baselines given assumed $(I,Q,U)$ maps.  The latter step is discussed in Section~\ref{sec-destripe}.  For the former, Equation~\ref{eq-sigmodel} is the basis of the least-squares fit for each pixel $p_0$ in the map. Summing over all detectors $i$, scans $j$, and samples $k$ that observe $p_0$, the associated $\chi^2$ of the estimated $(I_{p0}, Q_{p0}, U_{p0})$ is:

\begin{align}
\chi^2 &= \sum_{p_{i,j,k}=p_0} \frac{1}{\sigma^2(S_{i,j,k})} [S_{i,j,k} - b_{i,j} - \nonumber \\ &g_i(I_{p_0} + \epsilon_i [Q_{p_0} \cos 2(\theta_i-\phi_j) + U_{p_0} \sin 2(\theta_i-\phi_j)])]^2
~~~.
\label{eq-chi2}
\end{align}

Minimization of $\chi^2$ leads to coupled equations:

\begin{align} \label{eq-matrix}
\begin{pmatrix}
\sum\limits_{p=p_0}\frac{g^2_i}{\sigma^2(S_{i,j,k})} &
\sum\limits_{p=p_0}\frac{g^2_i\epsilon_i \cos 2(\theta_i-\phi_j)}{\sigma^2(S_{i,j,k})} &
\sum\limits_{p=p_0}\frac{g^2_i\epsilon_i \sin 2(\theta_i-\phi_j)}{\sigma^2(S_{i,j,k})} \\
\sum\limits_{p=p_0}\frac{g^2_i\epsilon_i \cos 2(\theta_i-\phi_j)}{\sigma^2(S_{i,j,k})} &
\sum\limits_{p=p_0}\frac{g^2_i\epsilon^2_i \cos^2 2(\theta_i-\phi_j)}{\sigma^2(S_{i,j,k})} &
\sum\limits_{p=p_0}\frac{g^2_i\epsilon^2_i \cos 2(\theta_i-\phi_j) \sin 2(\theta_i-\phi_j)}{\sigma^2(S_{i,j,k})} \\
\sum\limits_{p=p_0}\frac{g^2_i\epsilon_i \sin 2(\theta_i-\phi_j)}{\sigma^2(S_{i,j,k})} &
\sum\limits_{p=p_0}\frac{g^2_i\epsilon^2_i \cos 2(\theta_i-\phi_j) \sin 2(\theta_i-\phi_j)}{\sigma^2(S_{i,j,k})} &
\sum\limits_{p=p_0}\frac{g^2_i\epsilon^2_i \sin^2 2(\theta_i-\phi_j)}{\sigma^2(S_{i,j,k})} \\
\end{pmatrix}
\begin{pmatrix}
I_{p_0} \\
Q_{p_0} \\
U_{p_0} \\
\end{pmatrix}
= \nonumber \\
\begin{pmatrix}
\sum\limits_{p=p_0}\frac{g_i(S_{i,j,k}-b_{i,j})}{\sigma^2(S_{i,j,k})} \\
\sum\limits_{p=p_0}\frac{g_i\epsilon_i \cos 2(\theta_i-\phi_j)(S_{i,j,k}-b_{i,j})}{\sigma^2(S_{i,j,k})} \\
\sum\limits_{p=p_0}\frac{g_i\epsilon_i \sin 2(\theta_i-\phi_j)(S_{i,j,k}-b_{i,j})}{\sigma^2(S_{i,j,k})} \\
\end{pmatrix}
\end{align}

Some idealized cases are helpful for turning Equation~\ref{eq-matrix} into guidance for instrument design.  Consider a simplified situation with uniform detector noise $\sigma(S_{i,j,k})$, gain $g_i$, and polarization efficiency $\epsilon_i$; and where the observing method and map making allow the baselines $b_{i,j}$ to be measured and subtracted with negligible noise contribution.  As a first special case of that, assume that the focal plane layout and observing method produce a nearly random distribution of angles $\theta_i-\phi_j$.  Therefore, $<\cos 2(\theta_i-\phi_j)> \approx 0$, $<\cos^2 2(\theta_i-\phi_j)> \approx \frac{1}{2}$, etc., resulting in the following simplified solutions:
\begin{align}
I_{p0} &\approx g^{-1}<S_{i,j,k}>_{p=p_0}\label{eq-Iapprox}\\
Q_{p0} &\approx 2g^{-1}\epsilon^{-1}<\cos 2(\theta_i-\phi_j)S_{i,j,k}>_{p=p_0} \\
U_{p0} &\approx 2g^{-1}\epsilon^{-1}<\sin 2(\theta_i-\phi_j)S_{i,j,k}>_{p=p_0}
\end{align}
and associated uncertainties:
\begin{align}
\sigma(I_{p0}) &\approx g^{-1}N^{-1/2}_{p=p_0}\sigma(S_{i,j,k}) \\
\sigma(Q_{p0}) &\approx \sqrt{2}g^{-1}\epsilon^{-1}N^{-1/2}_{p=p_0}\sigma(S_{i,j,k}) \\
\sigma(U_{p0}) &\approx \sqrt{2}g^{-1}\epsilon^{-1}N^{-1/2}_{p=p_0}\sigma(S_{i,j,k})\label{eq-sigU}
\end{align}
i.e., $\sigma(Q_{p0}) \approx \sigma(U_{p0}) \approx \sqrt{2}\epsilon^{-1}\sigma(I_{p0})$.  $N_{p=p_0}$ is the number of observations of position $p_0$.

As a second special case of uniform $\sigma(S_{i,j,k})$, $g_i$, and $\epsilon_i$, consider an observation at fixed instrument rotation angle $\phi_j$ using detectors divided evenly among three polarization angles $\theta_i$ separated by 60$^\circ$ (or 120$^\circ$); this describes the intent of PPI (Section~\ref{sec-fp}).  It is straightforward to show that the Stokes vector solution and uncertainties are the same as in the random case (Equations~\ref{eq-Iapprox}-\ref{eq-sigU}).  Furthermore, this is also the case for four or more polarization orientations equally distributed over 180$^\circ$.  This means that there is no signal-to-noise disadvantage for using the minimum number of three angles.

The condition of the $3\times 3$ matrix on the left-hand side of Equation~\ref{eq-matrix} is a figure of merit for the quality of the $(I,Q,U)$ measurement in the pixel of interest\cite{Kurki-Suonio:2009}.  A larger determinant means a lower uncertainty in the solution.  In this paper, we use the term ``Stokes hits'' for the cube root of the determinant of the $3\times 3$ matrix when $\sigma^2(S_{i,j,k})$ and $g$ are set to unity.  For the simplified case in Equations~\ref{eq-Iapprox}-\ref{eq-sigU}, we can recognize $N_{p=p_0}$ as the ``hits'' for total intensity measurement.  In that same case, the ``Stokes hits'' works out to $2^{-2/3}\epsilon^{4/3}N_{p=p_0}$, up to 63\% of the total intensity hits.

\subsection{Map Destriping and Detector 1/f Noise\label{sec-destripe}}

For the observing mode that will be discussed and simulated in Section~\ref{sec-sim}, the observation sequence is clearly divided into discrete scans, and it is natural to divide the detector timestreams into segments (one per scan) and to fit a detector baseline for each segment.  With the condition that one scan corresponds to one pass over the observed source, and the additional condition that only a constant baseline is fit per detector and per scan, then the only filtering of spatial modes that occurs is removal of a constant value over the map.  The least-squares solution for the baselines, given the estimate of the Stokes parameter maps, is straightforward:

\begin{equation} \label{eq-baselines}
b_{i,j} = (\sum_k \frac{1}{\sigma^2(S_{i,j,k})})^{-1} \sum_k \frac{S_{i,j,k} - g_i(I_{p(i,j,k)} + \epsilon_i [Q_{p(i,j,k)} \cos 2(\theta_i-\phi_j) + U_{p(i,j,k)} \sin 2(\theta_i-\phi_j)])}{\sigma^2(S_{i,j,k})}
~~~.
\end{equation}

\begin{figure}
\begin{center}
\includegraphics[width=\linewidth]{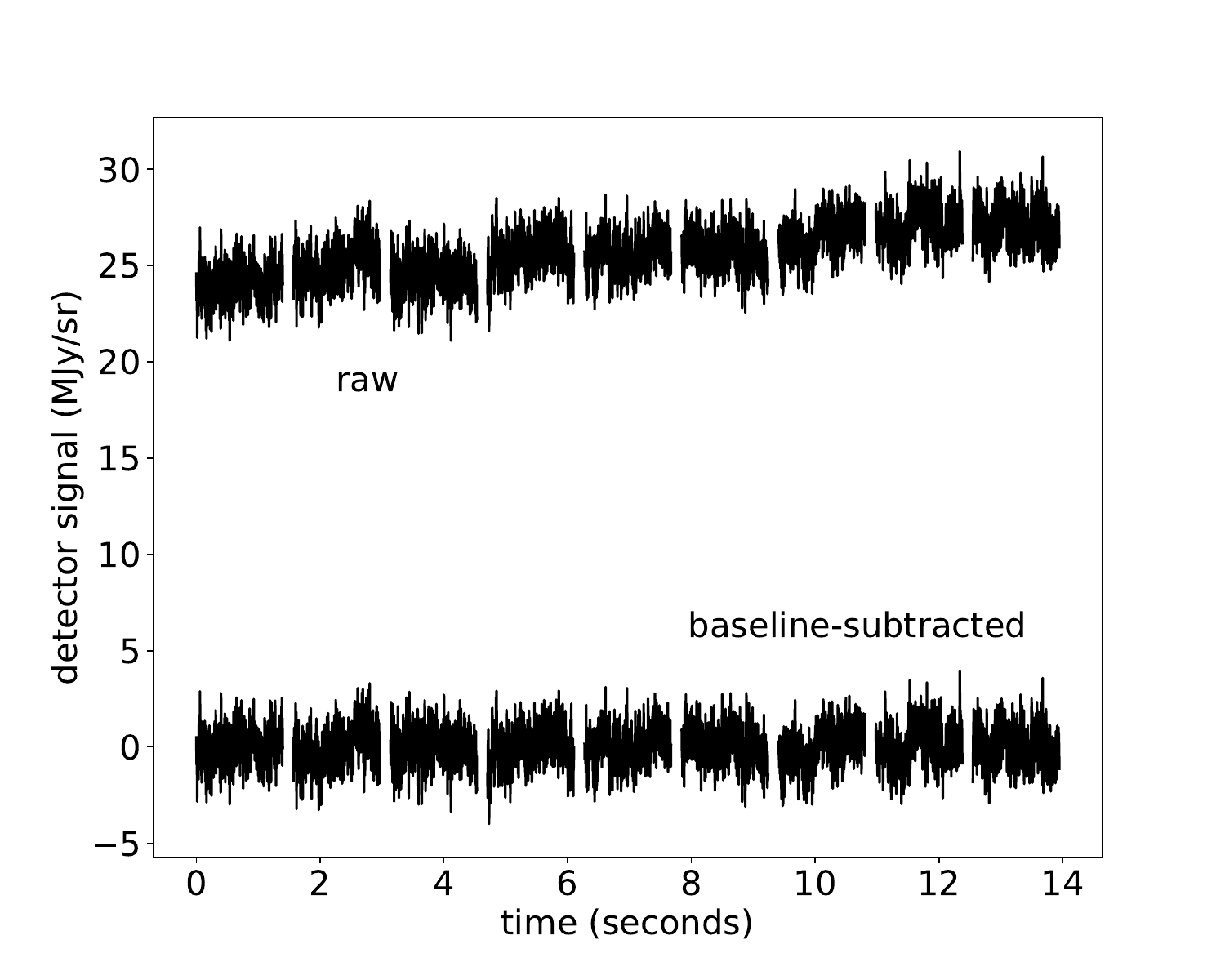}
\caption{\label{fig:timestream}  Illustration of detector $1/f$ noise model (Section~\ref{sec-fp}) and baseline subtraction by segments.  The upper curve shows a noise model realization for one detector, for the first 14 seconds of an observation.  The detectors are sampled at $f_{samp}=$350\,Hz.
The gaps in the curve are the turn-around periods between scan segments, for which the detector samples are not used in the map making.  The lower curve shows the detector signal after subtraction of the baselines (one for each segment) found from the iterative map solution.
}
\end{center}
\end{figure}

Subtracting baselines on each scan segment separately is an effective way to remove $1/f$ noise from the detector signals on timescales longer than the scan duration (Figure~\ref{fig:timestream}; e.g., \citenum{Kurki-Suonio:2009}).  This paper concentrates on $1/f$ noise uncorrelated among detectors.  PRIMA is being designed to address potential correlated $1/f$ noise in other ways, such as using detector ``blind tones''\cite{Baselmans2017} to subtract drift in the electronics and thermal control and monitoring.

Expanding on the mention in Section~\ref{sec-map-intro} of arbitrary value of the zero level in the output maps, we can now describe this in terms of the signal model.  If one expands Equation~\ref{eq-chi2} to sum over all of the positions \{p\} in the map, then attempts to solve for all $(I,Q,U)$ and $\{b\}$ simultaneously, then one encounters a singular matrix and degeneracy among certain parameter combinations.  Specifically, if the substitutions $b_{i,j} \rightarrow b_{i,j} - g_i \Delta I$ and $I_{p} \rightarrow I_{p} + \Delta I$ are made for the solutions, then the value of $\chi^2$ is unchanged, i.e., changing $I$ by a constant amount over the full map (with a corresponding change in baselines) does not degrade the goodness of fit.  Fortunately, the degeneracy exists only for the spatially-constant term.  In addition to $I$, the degeneracy exists for the two constant terms in polarization also; for example, the substitutions $b_{i,j} \rightarrow b_{i,j} - g_i \epsilon_i \cos 2(\theta_i-\phi_j) \Delta Q$ and $Q_{p} \rightarrow Q_{p} + \Delta Q$ also do not change the value of $\chi^2$.

\section{Polarimetry Simulation\label{sec-sim}}

To demonstrate the polarization-mapping capability of PPI, and to quantify the effect of detector $1/f$ noise, we simulated an observation of a PRIMA science target and carried out the map-making process described in Section~\ref{sec-mapping}.

\subsection{Focal Plane Detector Array\label{sec-fp}}

\begin{figure}
\begin{center}
\includegraphics[width=\linewidth]{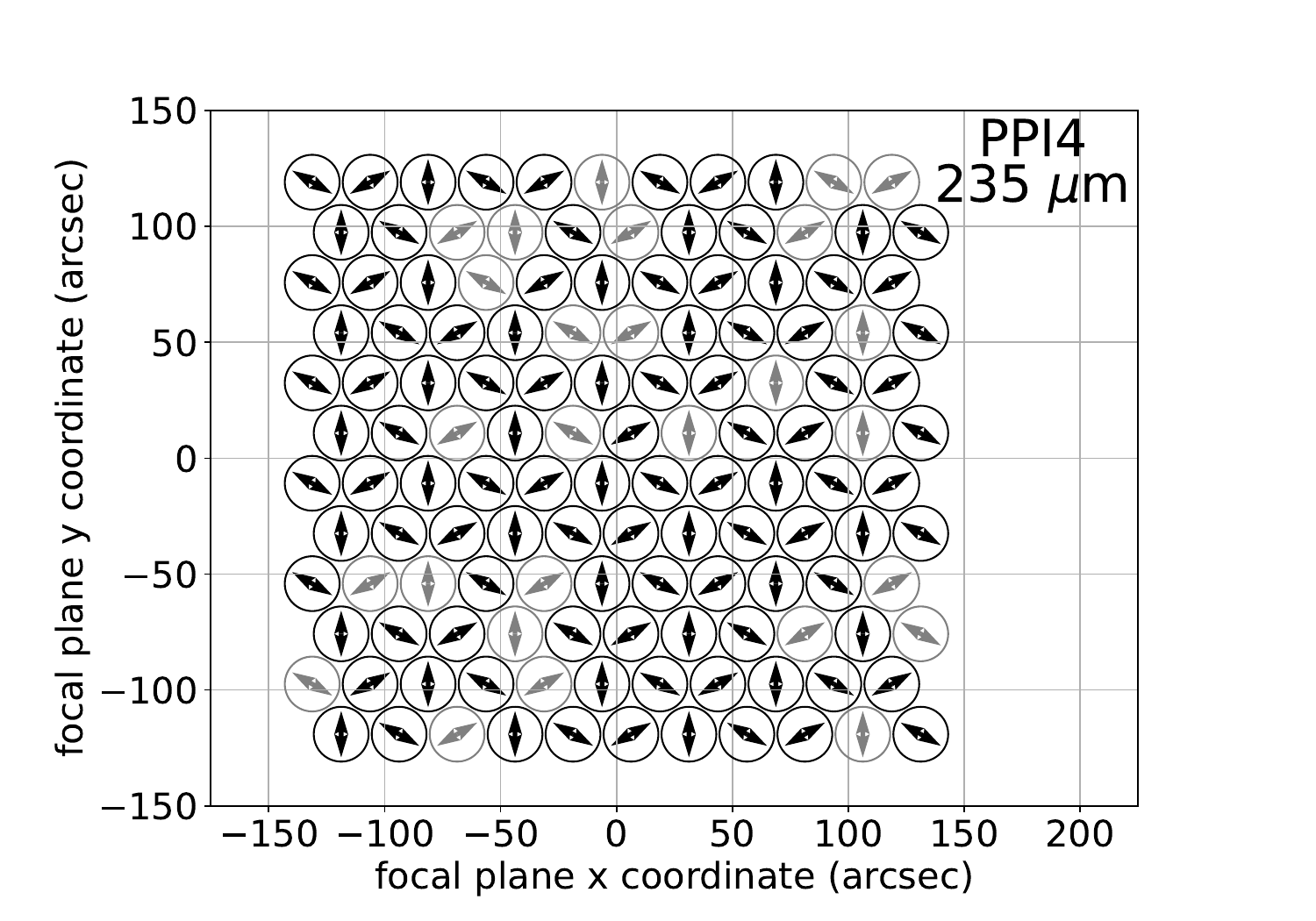}
\caption{\label{fig:focal_plane} Focal plane layout of PPI4.  132 microlens-fed kinetic inductance detectors are arranged in a hexagonal-close-packed configuration with nearest-neighbor spacing of 25$^{\prime\prime}$\ on the sky.  The arrows show the direction of sensing linear polarization for each detector, dictated by the KID antenna design.  Three polarization angles are distributed evenly through the focal plane.  The detectors shown in gray were omitted (by random number generation) from the mapping simulations to capture the pessimistic detector yield.  PPI1 has 1116 detectors arranged in a similar pattern with 10$^{\prime\prime}$\ nearest-neighbor separation.}
\end{center}
\end{figure}

For the simulations in this paper, we model the shortest- and longest-wavelength bands of the PRIMAger Polarimetric Imager: PPI1 with bandpass center at $\lambda = 92$~$\mu$m\ and PPI4 at $\lambda = 235$~$\mu$m.  Each detector pixel consists of a hemispherical microlens which concentrates the intercepted power onto an antenna designed to be sensitive in a single linear polarization \cite{Baselmans2022}.  The direction of polarization is varied in a regular way over the detector array such that the full array is sensitive to three angles of polarization separated by 120$^\circ$ with approximately equal number of detectors allocated for each direction.  The power collected by each antenna is absorbed in a microwave resonator circuit with resonance frequency dependent on the input power.  The PPI1 detectors are arranged on a hexagonal grid with nearest-neighbor separation of 10$^{\prime\prime}$\ in the far field, and the estimated beam size is 11$^{\prime\prime}$\ Full-Width-Half-Maximum (FWHM).  For PPI4 (Figure~\ref{fig:focal_plane}), the detector separation is 25$^{\prime\prime}$ , and the estimated beam size is 27$^{\prime\prime}$\ FWHM.

Noise in the simulation arises from two sources:  the detector and photon statistics.  To compute the latter, the specific intensity of the source (and background) $I_\nu$ is converted to power as follows:
\begin{equation}
P_{phot} = \frac{n_{pol}}{2}\eta (\frac{A\Omega}{\lambda^2})\lambda^2\frac{\nu}{R}I_\nu
\end{equation}
and the Noise Equivalent Power (NEP, in ${\rm W}\,{\rm Hz}^{-1/2}$) is calculated as (see, e.g., \citenum{Mather:1982}):
\begin{equation}
NEP_{phot} = \sqrt{2P_{phot}h\nu}
~~~.
\end{equation}
In the prior equations, $n_{pol}$ is the number of polarization states detected (1 for PPI), $\eta$ is optical efficiency including the detector, $A\Omega$ is the etendue, $\lambda$ and $\nu$ are the wavelength and frequency, $P_{phot}$ is the power absorbed in the detector, and $h$ is the Planck constant; see also Table~\ref{tbl-instpar}.  For the simulation, the photon coherence term has not been included since it makes a negligible contribution (fractional increase $<10^{-4}$ for NGC\,6946 and PPI bands).  The noise intrinsic to the detector is assumed to be independent of the photon power for the intensity range in this paper
so a fixed $NEP_{det}$ adds in quadrature with $NEP_{phot}$ to give $NEP_{tot}$.  (Although this assumption is not accurate in detail for KIDs \cite{Baselmans2022}, it is a reasonable approximation for the purpose of this simulation.)  The Noise Equivalent Intensity (NEI) for a given detector is defined as follows:
\begin{equation}
NEI = NEP_{tot} / [\frac{n_{pol}}{2}\eta (\frac{A\Omega}{\lambda^2})\lambda^2\frac{\nu}{R}]
\end{equation}
For low-background lines of sight with $I_\nu$ = 7\,MJy/sr, and the ``modeled'' parameter assumptions in Table~\ref{tbl-instpar}, the NEI for PPI1 is 0.182\,MJy/sr/Hz$^{1/2}$, and for PPI4 is 0.071\,MJy/sr/Hz$^{1/2}$.  The uncertainty in a measurement of specific intensity $I_\nu$ is related to $NEI$ and integration time $T$ as follows:  $\sigma(I_\nu) = \frac{NEI}{\sqrt{2T}}$ (e.g., \citenum{Rieke:1996}).

\begin{sidewaystable}[ht]
\caption{Predicted and modeled (pessimistic) PPI instrument parameters.} 
\small
\label{tbl-instpar}
\begin{center}       
\begin{tabular}{lccccl} 
\hline
\rule[-1ex]{0pt}{3.5ex} Parameter & PPI1 predicted & PPI1 modeled & PPI4 predicted & PPI4 modeled & Unit \\
\hline\hline
\rule[-1ex]{0pt}{3.5ex} wavelength $\lambda$ & 92 & 92 & 235 & 235 & $\mu$m \\
\rule[-1ex]{0pt}{3.5ex} $NEP_{det}$ (white comp.) & $\leq3.0\times 10^{-19}$ & $6.0\times 10^{-19}$ & $\leq3.0\times 10^{-19}$ & $6.0\times 10^{-19}$ & ${\rm W}\ {\rm Hz}^{-1/2}$ \\
\rule[-1ex]{0pt}{3.5ex} $NEP_{det}$ ($1/f$ comp.) & $\leq6.0\times 10^{-20}(\frac{f}{10 {\rm Hz}})^{-0.6}$ & $6.0\times 10^{-19}(\frac{f}{10 {\rm Hz}})^{-0.6}$ & $\leq6.0\times 10^{-20}(\frac{f}{10 {\rm Hz}})^{-0.6}$ & $6.0\times 10^{-19}(\frac{f}{10 {\rm Hz}})^{-0.6}$ & ${\rm W}\ {\rm Hz}^{-1/2}$ \\
\rule[-1ex]{0pt}{3.5ex} optical efficiency $\eta$ & 0.29 & 0.15 & 0.29 & 0.15 & \\
\rule[-1ex]{0pt}{3.5ex} detector $A\Omega/\lambda^2$ & 0.74 & 0.74 & 0.74 & 0.74 & \\
\rule[-1ex]{0pt}{3.5ex} polarization efficiency $\epsilon$ & $>0.99$ & 0.99 & $>0.99$ & 0.99 \\
\rule[-1ex]{0pt}{3.5ex} NEI (7\,MJy/sr bgnd.) & 0.077 & 0.182 & 0.030 & 0.071 & MJy/sr/Hz$^{1/2}$ \\
\rule[-1ex]{0pt}{3.5ex} sample yield$^a$ & $\geq0.92$ & 0.90 & $\geq0.92$ & 0.90 & \\
\rule[-1ex]{0pt}{3.5ex} detector yield & $\geq0.80$ & 0.80 & $\geq0.80$ & 0.80 & \\
\hline
\end{tabular}
\end{center}
$^a$Due to cosmic rays
\end{sidewaystable}


The simulation first generates ``photon only'' detector timestreams by looking up $(I,Q,U)$ for the relevant positions on the sky and applying Equation~\ref{eq-sigmodel}.  (PRIMA is being designed such that thermal emission from the optics and stray light are insignificant contributors to photon noise compared to the astrophysical background.)  In the same loop through detectors and samples, photon noise is applied by adding a random number drawn from a gaussian distribution with mean 0 and standard deviation ($\frac{f_{samp}}{2})NEP_{phot}$, where $\frac{f_{samp}}{2}$ is the bandwidth of the detector samples.

The detector noise is modeled with a white noise component and a $1/f^n$ component.  For each detector and for the full period of time required to complete one repetition of the scan pattern, a long timestream is modeled using the prescription in the prior sentence for the power spectral density, along with randomized phases.  Each repetition of the scan pattern is modeled independently.  The detector noise timestream is added to the ``photon only'' timestream, which is then added to a large, random baseline constant over the repetition.  Measurements of prototype PRIMAger KIDs (similar to those reported by \citenum{Baselmans2022}) indicate a $\sim\! 1/f^{0.6}$ spectrum; in the simulations, the amplitude of this component has been exaggerated by a factor of 10 to examine robustness of the observation approach to noise of this type.

For simplicity, the simulation uses uniform weighting -- constant $\sigma^2(S_{i,j,k})$ -- for all detectors contributing to a map pixel.  This approximates the case of a focal plane with a small distribution of detector noise properties.  Since we are not yet in the situation of correcting for a distribution of detector gains and polarization efficiencies, we set $g_i = 1$ (until Section~\ref{sec-gain}) and $\epsilon_i = 0.99$ for the simulations.

To simulate a focal plane with imperfect yield, 20\% of the detectors haven been omitted from the simulation randomly (Figure~\ref{fig:focal_plane}).  To simulate the data loss from cosmic rays, 10\% of the detector samples are omitted randomly.  This is a conservative estimate, as we expect losses of between 2\% and 8\% based on scaling from laboratory measurements of representative KIDs \cite{Kane2024}, and assuming simple masking of the impacted timestream.  Cosmic ray processing is done on board, with a base sample rate of 10\,kHz, prior to rebinning to 350~Hz for downlink.  The data loss may be reduced by subtracting the tails of the events, as was done with Planck \cite{Catalano2014}, but in the on-board electronics in the case of PRIMA. We found from simulation that the only significant impact of the omission of samples from the mapping is the loss of integration time.

\subsection{Simulated Science Target\label{sec-target}}

\begin{figure*}
\begin{center}
\includegraphics[width=\textwidth]{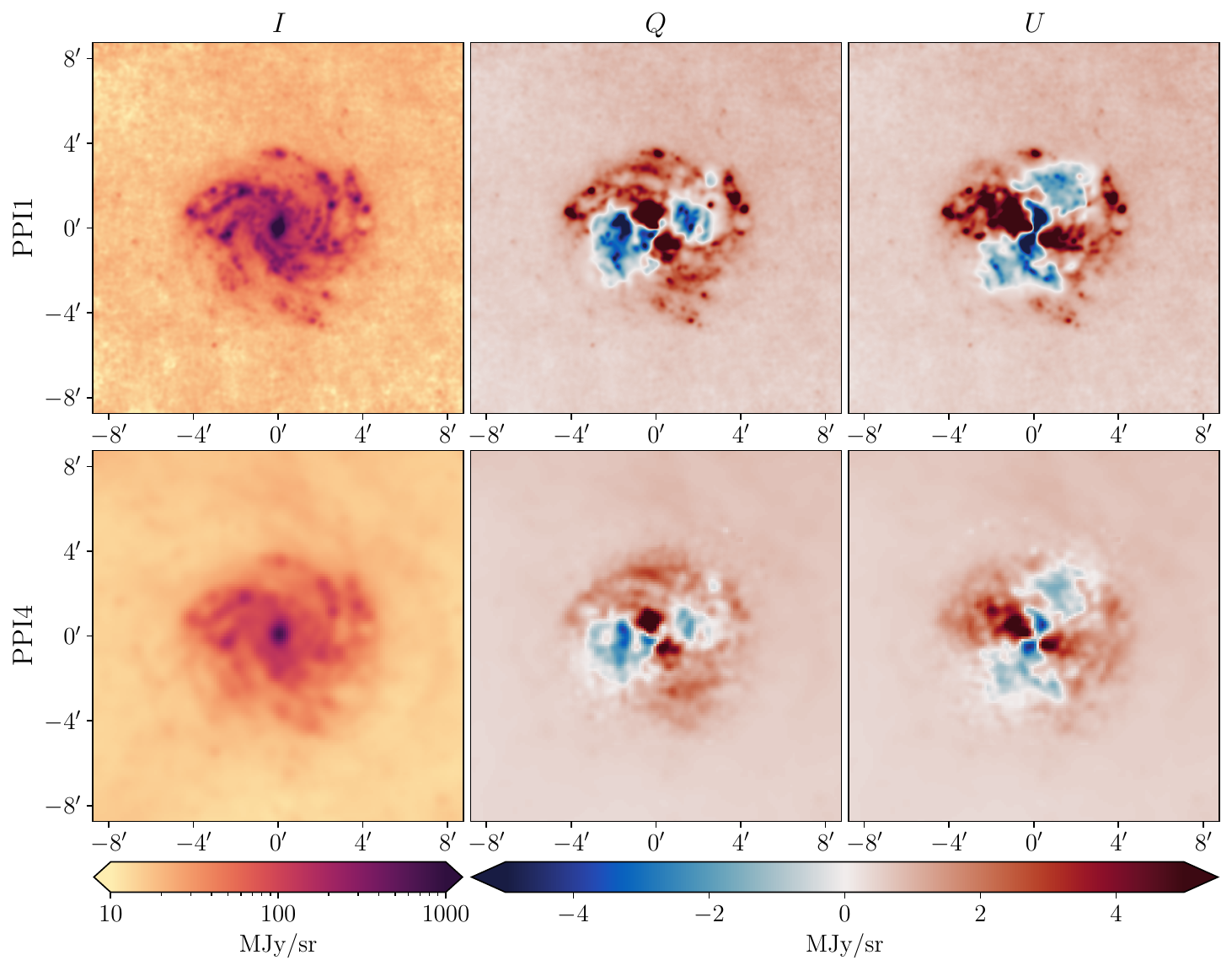}
\caption{\label{fig:truth_maps} Input ``truth'' $I$ (left), $Q$ (middle), and $U$ (right) maps for PPI1 (top) and PPI4 (bottom) derived from Herschel and VLA observations as described in Section~\ref{sec-target}.  By choice of the cirrus model, the $Q$ and $U$ ``foreground'' emission at the periphery of the maps is positive.
}
\end{center}
\end{figure*}

As an illustrative test case of our polarimetric mapmaking techniques, we consider a mock observation of the nearby face-on spiral galaxy NGC\,6946. Its FIR size of 100 arcmin$^2$ \cite{Aniano:2020} is well-matched to mapping with PPI while a wealth of ancillary data, including FIR and radio imaging and polarimetry, can inform our signal model. Although we hew closely to available data where possible, for the purposes of this work we prioritize a signal model that best illustrates PPI mapmaking capabilities and limitations rather than a strict forecast; we note our simplifying assumptions below.

Our PPI1 simulations are based on 100\,$\mu$m observations with the Herschel/PACS instrument (ESA Herschel Science Archive, OBSIDs 1342191947-50). These data have an angular resolution of 7$^{\prime\prime}$\ and a pixel size of 1.6$^{\prime\prime}$. We first smooth the map to a PPI1 resolution of 11$^{\prime\prime}$\ with a pixel size of 3$^{\prime\prime}$\ using the astropy \texttt{reproject} package. We extract a 350$\times$350 pixel region centered on the nucleus for our fiducial Stokes $I$ map to avoid noise at the edges of the PACS map. For simplicity, we adopt the flux densities as-is, i.e., without accounting for the slightly shorter wavelength of PPI1.

Our PPI4 simulations are based on 250\,$\mu$m observations with the Herschel/SPIRE instrument (ESA Herschel Science Archive, OBSIDs 1342183046-53, 1342183364, 1342183366, 1342188786, 1342266676). Unlike the PACS map, the SPIRE map has an overall zero level of approximately 12.2\,MJy\,sr$^{-1}$, which we remove for our initial processing. The SPIRE map has an angular resolution of 17$^{\prime\prime}$\ and has 6$^{\prime\prime}$\ pixels. We smooth to a PPI4 resolution of 27$^{\prime\prime}$\ and a pixel size of 7$^{\prime\prime}$. We extract a 400$\times$400 pixel region centered on the nucleus for our fiducial Stokes $I$ map. As with the PPI1 map, we do not attempt to correct for the small wavelength difference between PPI4 and Herschel and so use the measured flux densities as-is.

For the Stokes $Q$ and $U$ maps, we start from 6\,cm observations\cite{Borlaff2023} of NGC\,6946 with the Very Large Array (VLA)
(FIR $Q$ and $U$ maps of NGC\,6946 were made with SOFIA \cite{Borlaff2023}, but these maps have lower signal-to-noise ratio over less area of the galaxy than the radio data and so are less suitable for simulation purposes.)
While the polarized intensity of the synchrotron emission measured at 6\,cm is unlikely to be strongly correlated with dust polarization at shorter wavelengths, the orientation of the galaxy-scale magnetic field as probed by these tracers should be similar. Therefore, we adopt the polarization angles of the radio data. We make the additional assumption that the synchrotron polarization fraction is perfectly correlated with the dust polarization fraction. While unlikely to be true in practice, this provides a straightforward means of inducing polarization fraction variations across the map. We reduce the radio polarization fraction by a factor of three to account for the fact that the intrinsic dust polarization fraction is less than that of synchrotron emission. The resulting distribution of polarization fractions is broadly consistent with FIR measurements of nearby galaxies \cite{Lopez-Rodriguez:2022}.

Explicitly, the PPI1 and PPI4 $Q$ and $U$ maps are constructed from the radio $Q_R$ and $U_R$ maps, the PPI1 and PPI4 $I$ maps, and the radio $I_R$ map as

\begin{align}
    Q_{\rm PPIX} &= 0.33 I_{\rm PPIX} \frac{Q_R}{I_R} \\
    U_{\rm PPIX} &= 0.33 I_{\rm PPIX} \frac{U_R}{I_R}
\end{align}
where X denotes either 1 or 4.

While the radio data provide a constraint on the magnetic field morphology across the disk of NGC\,6946, they do not probe the foreground Galactic cirrus that is evident in both Herschel maps. To account for the cirrus, we first construct an ansatz cirrus polarization fraction map that smoothly varies from 0\% in the southeast corner of the map to a value approaching 10\% in the northwest corner, consistent with observed values of Milky Way cirrus across the sky \cite{planck2014-XIX}. This variation is not necessarily physical nor based on data (e.g., from Planck), but rather is a means of introducing variable cirrus polarization fractions into the simulations. We assume that the cirrus has equal power in $Q$ and $U$ to construct the final cirrus $Q$ and $U$ maps from our cirrus polarization fraction map and the Herschel $I$ maps. We employ an apodized mask to smoothly transition from the cirrus $Q$ and $U$ maps to the galaxy $Q$ and $U$ maps based on the radio data.

Finally, we reintroduce a zero level back into both $I$ maps. As Galactic dust has comparable brightness at 92 and 235\,$\mu$m (e.g., \citenum{Dwek:1997}), we employ a value of 12.2\,MJy\,sr$^{-1}$ at both wavelengths. Since this zero level is likely dominated by Galactic cirrus, with NGC\,6946 being just 11.7$^\circ$ off the Galactic plane, we add this zero level to the $Q$ and $U$ maps using our cirrus polarization fraction map assuming equal power in $Q$ and $U$.

The final $I$, $Q$, and $U$ maps for PPI1 and PPI4 are presented in Figure~\ref{fig:truth_maps}. These constitute the ``ground truth'' for our simulated observations.

\subsection{Scan Pattern\label{sec-scan}}

To guide the choice of the scan pattern, we assume that the PPI detectors are sampled at $f_{samp} = 350$\,Hz and have a useful bandwidth of 175\,Hz.  At the PPI shortest wavelength, $\frac{\lambda}{2D}$ Nyquist sampling of the sky is 5.2$^{\prime\prime}$, allowing a scan rate as high as 5.2$^{\prime\prime}$ $\times$ 350\,Hz = 1820$^{\prime\prime}$ /sec.  For the simulations in this study, we use a scan rate of 500$^{\prime\prime}$/sec.

PRIMA observations of science targets that are several arcminutes across, such as NGC\,6946, can take advantage of the BSM.  The BSM provides large scan rates along with large acceleration for time-efficient scan ``turn-arounds.''  The PRIMA optics and BSM allow regions of 24$^\prime$ $\times$ 42$^\prime$\ to be mapped without moving the bulk telescope and spacecraft.

\begin{figure}
\begin{center}
\includegraphics[width=\linewidth]{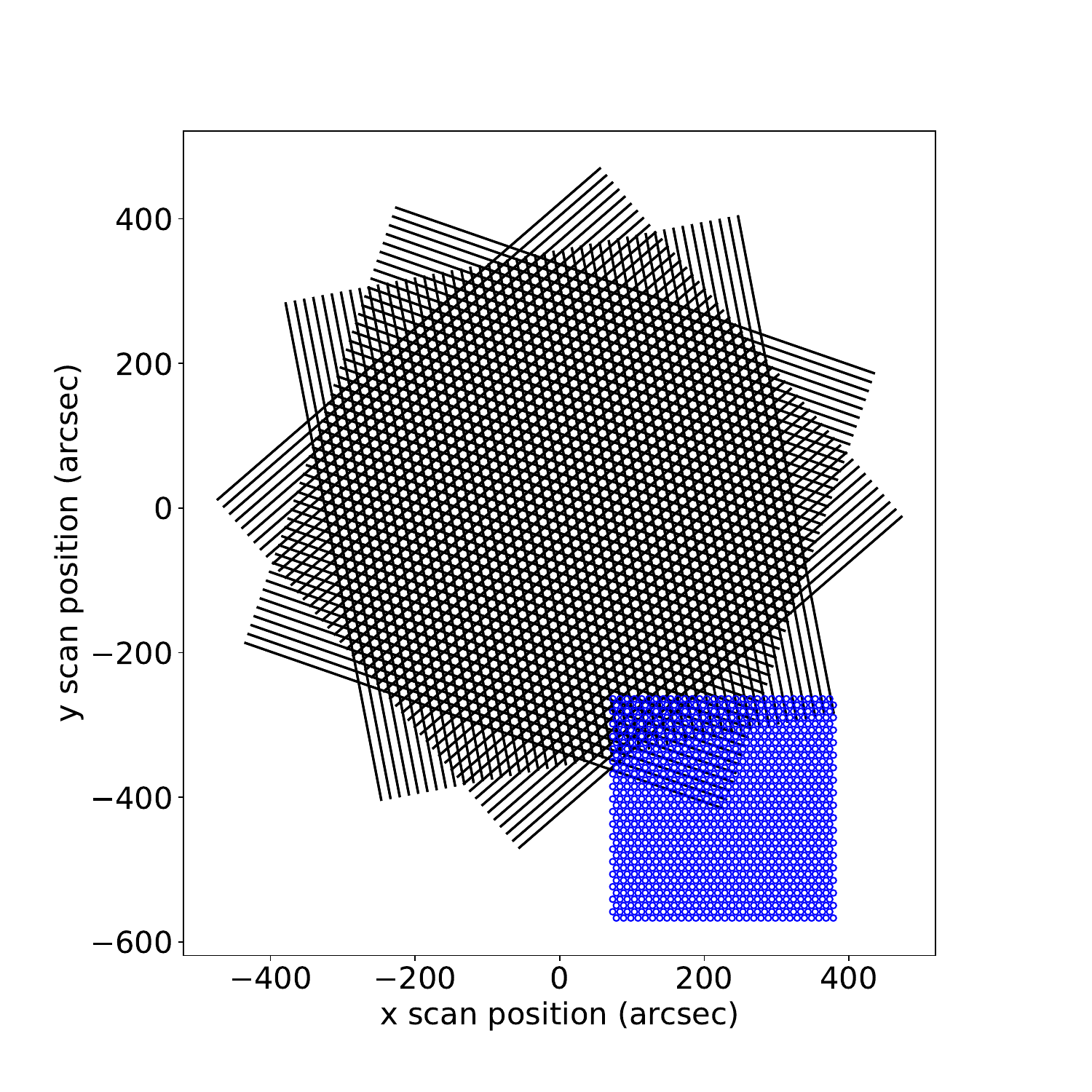}
\caption{\label{fig:scan_pattern} Scan pattern used in the simulation.  Each of the $3\times 50$ black lines shows the trajectory of the center of the detector array on a Cartesian sky coordinate system.  For scale, the $31\times 36$ layout of the PPI1 detector array is shown in blue at one of the extreme points of the scan.  The observation is 1.5 hour in duration, and rotation of the field is negligible over that time given PRIMA's L2 orbit.  The scan pattern fits within a diameter of 16$^\prime$ .  Adding the footprint of the detector array, the observed area is within a 21$^\prime$\ $\times$ 21$^\prime$\ rectangular region.}
\end{center}
\end{figure}

The scan pattern simulated for this study is shown in Figure~\ref{fig:scan_pattern}.  Three scan axes are used, matching the three-fold symmetry of the focal plane.  The exact directions were not found to be critical for successfully recovering the source Stokes vectors; however, the scan direction, along with the spacing of the raster lines, affects the uniformity of coverage.  The simulation uses scan directions at a 19.1$^\circ$ angle from the focal plane symmetry axes.  Each raster line is 700$^{\prime\prime}$\ long (lasting 1.40 seconds) and is covered in a back-and-forth way.  The spacing of the raster lines is 13$^{\prime\prime}$.  The simulation assumes that the BSM reverses the scan direction and (on half of the occasions) moves to the next raster line within 0.14 seconds, requiring an angular acceleration of the far-field beam of 7100 $^{\prime\prime}$/sec$^2$:  within the design specifications of the BSM.  There are $3\times 50\times 2 = 300$ total scan segments in the scan pattern, and the pattern is repeated 12 times in the simulation, so the total observation duration is 1.54 hour.

\subsection{Simulation Results} \label{sec-sim-results}

Simulations were performed for multiple noise realizations of PPI1 and PPI4.  Each realization used the scan pattern from Section~\ref{sec-scan} and Figure~\ref{fig:scan_pattern}, carried out for the 10 repetitions.  The target was ``observed'' in each each band independently (although an optimized scan pattern could be enlarged to observe multiple bands simultaneously and more efficiently).  From the simulated detector timestreams, $I$, $Q$, and $U$ maps were generated according to Equation~\ref{eq-matrix}, iterated with solution for the detector baselines (Equation~\ref{eq-baselines}). For convenience, the output map pixel scale matched the input (Section~\ref{sec-target}).  Convergence of the iteration is driven by the settling of the adjusted baselines.  For our choice to simulate baselines much larger than the astrophysical signal and $1/f$ noise on long timescales, convergence was sped up (with no observable impact to the final maps) by first subtracting the median signal from each detector for each scan segment prior to starting the iteration. 20 iterations were sufficient for convergence, judged by monitoring of $\chi^2$ (Equation~\ref{eq-chi2}) and the larger-scale patterns in the differences of output maps and input ``truth'' maps.

\begin{figure*}
\begin{center}
\includegraphics[width=\textwidth]{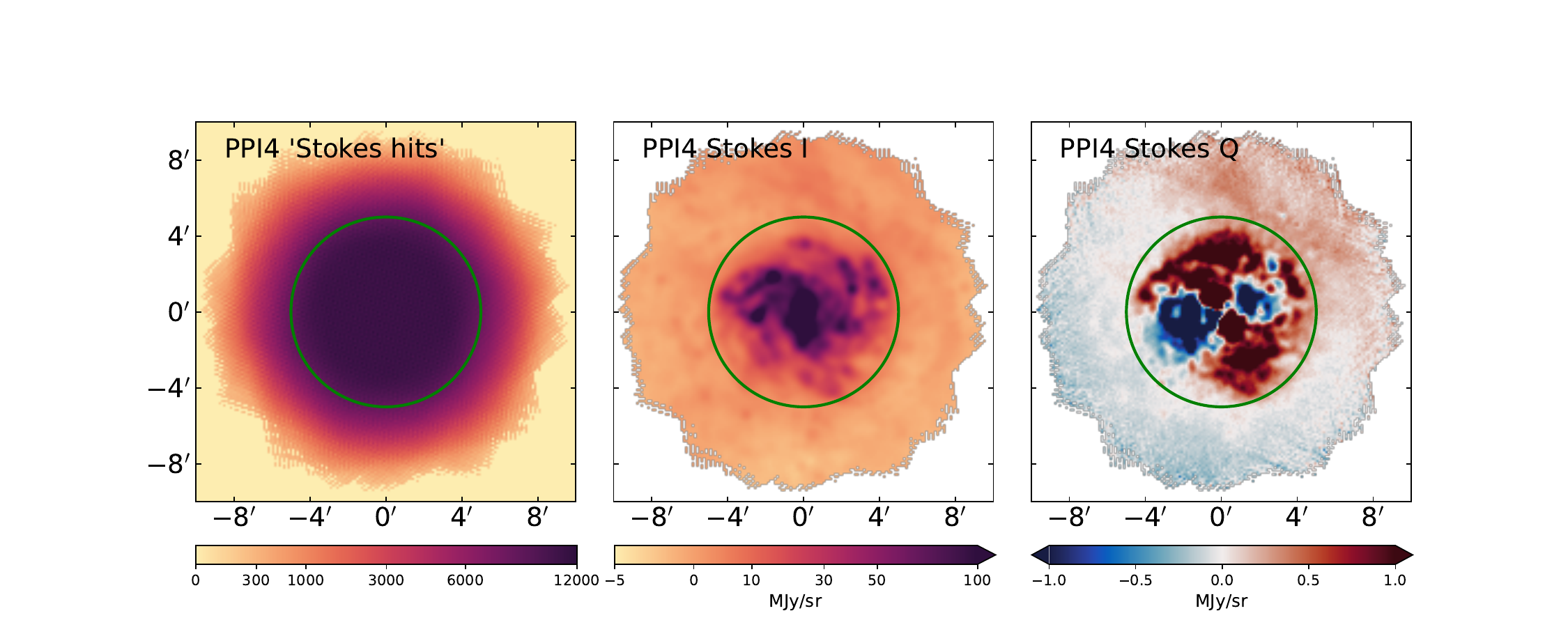}
\caption{\label{fig:PPI4_sim} A selection of the outputs from one observation realization with PPI4.  The left panel shows a map of ``Stokes hits'' (Section~\ref{sec-solution}).  The 10$^\prime$\ diameter science analysis region is shown with a green circle on all panels.  The Stokes hits has a peak value of 11,860 and a minimum value of 7570 within the 10$^\prime$\ diameter, for the 7$^{\prime\prime}$\ pixel scale used.  The $I$ and $Q$ images in the middle and right panels are displayed with saturated color scales to show low-level features.  The faint striping at the edges of the $Q$ map is a signature of the detector $1/f$ noise.}
\end{center}
\end{figure*}

\underline{PPI4 Simulation Results.}  Example output maps from the simulations are shown in Figure~\ref{fig:PPI4_sim}.  For PPI4, the median Stokes hits in the central 10$^\prime$\ diameter (left panel) is 10,620 per 7$^{\prime\prime}$\ pixel, which is 61\% of the median total intensity hits of 17,270 (49 seconds observing time per map pixel).  The ratio matches the expectation from Section~\ref{sec-solution} for good polarization angle coverage.  The reconstructed Stokes $I$ (center panel), $Q$ (right panel), and $U$ are mapped with high signal-to-noise in the central 10$^\prime$\ diameter.

\begin{figure*}
\begin{center}
\includegraphics[width=\textwidth]{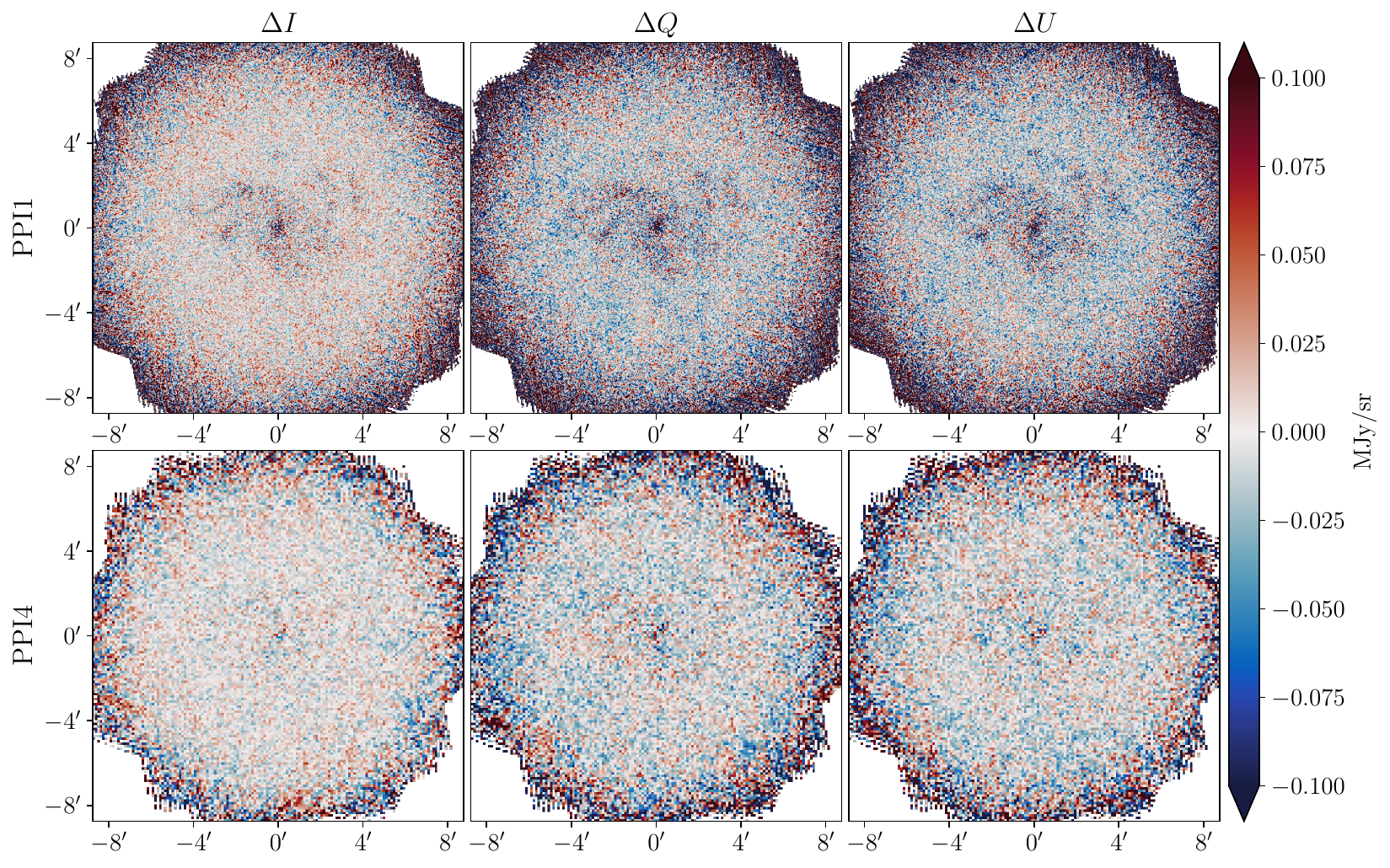}
\caption{Difference between the ``observed'' map and the input signal map in each of $I$ (left), $Q$ (middle), and $U$ (right) for both PPI1 (top) and PPI4 (bottom). All maps have had zero levels removed as described in Section~\ref{sec-analysis}. There is little structure in the map aside from an imprint of the galaxy where the pixel value scatter is dominated by photon noise and near the map edges where the hit count is low.\label{fig:residuals}}
\end{center}
\end{figure*}

The sensitivity of the maps can be quantified by taking differences of the input and output (1.5 hour observation) maps (Figure~\ref{fig:residuals}).  Note that we have applied a zero level correction to all maps for this comparison (as discussed further in Section~\ref{sec-analysis}). For PPI4, the standard deviation of the pixels in the central 10$^\prime$\ diameter of the $I$ difference is 0.0140\,MJy/sr.  This is 1.35$\times$ the white-noise-only expectation of 0.0104\,MJy/sr based on the median $I$ of 38\,MJy/sr, 49 seconds of integration, and the equations in Section~\ref{sec-fp}.  The standard deviation of the $Q$ (or $U$) difference in the 10$^\prime$\ diameter is 0.0200\,MJy/sr, also $1.35\times$ the white-noise prediction.  This excess noise is due to the $1/f^n$ noise not included in the prediction.  Statistics of the multiple observational realizations indicate similar values for the sensitivity in the maps.

The photon noise from NGC\,6946 and the Galactic foreground in its vicinity partially ``hide'' the detector $1/f^n$ noise.  To quantify this, we also performed a simulation with an input map which had spatially-uniform $I$ = 7\,MJy/sr and $Q = U = 0$.  Otherwise, the instrument and observational parameters were the same as for NGC\,6946.  For the 1.5 hour observation, we found a per-pixel sensitivity of 0.0110\,MJy/sr (1$\sigma$) in Stokes $I$ and 0.0159\,MJy/sr (1$\sigma$) in Stokes $Q$ or $U$, both 1.5$\times$ the theoretical expectation with white noise only.

\underline{PPI1 Simulation Results.}  For PPI1, the median Stokes hits in the central 10$^\prime$\ diameter is 16,270 per 3$^{\prime\prime}$\ pixel, which is 62\% of the median total intensity hits of 26,200 (75 seconds).  Based on differences of output and input maps, the per-pixel sensitivity for 1.5 hours observing, within the central 10$^\prime$\ diameter of NGC\,6946, is 0.0309\,MJy/sr (1$\sigma$) in Stokes $I$ and 0.0443\,MJy/sr (1$\sigma$) in Stokes $Q$ or $U$.  This is 1.37$\times$ the white-noise-only expectations of 0.0225 and 0.0321\,MJy/sr, respectively, based on the median $I$ of 44\,MJy/sr and 75 seconds of integration.

\begin{figure}
\begin{center}
\includegraphics[width=\linewidth]{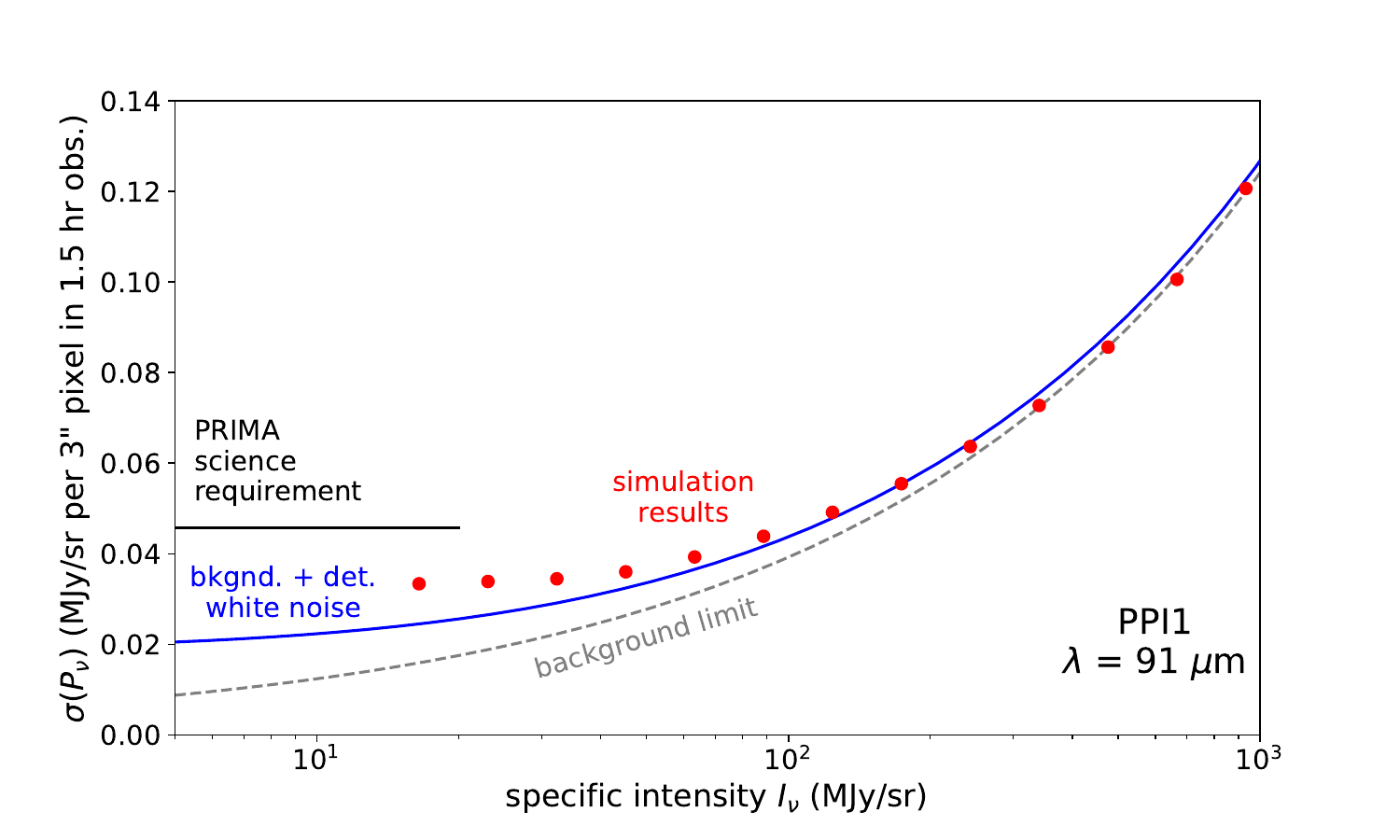}
\caption{\label{fig:sens_vs_I} Comparison of sensitivity measured from simulated PPI1 observations and two limiting cases.  The sensitivity is shown for polarized intensity $P_\nu = \sqrt{Q_\nu^2 + U_\nu^2}$, and the uncertainty is estimated as $\sigma(P_\nu) \approx (\sigma(Q_\nu) + \sigma(U_\nu))/2$.  The sensitivity is measured per pixel for a 1.5 hour observation resulting in 75 sec average integration time per pixel.  The gray dashed line shows the theoretical sensitivity (``background limit'') for an instrument with no detector noise and an end-to-end optical efficiency of 0.15.  The solid blue curve shows sensitivity including the modeled pessimistic detector white noise.  The red points are determined from the simulated observations, which include the modeled pessimistic detector $1/f^n$ noise.  For each intensity bin, the standard deviation is computed for the difference of output and input maps.  The sensitivity is compared to the PRIMA polarimetry science requirement of 5$\sigma$ = 0.030 MJy/sr in 2 hr, converted from PPI4 beam area to 3$^{\prime\prime}$\ pixels as described in Section~\ref{sec-analysis}.  The simulated map sensitivity is better than the requirement, despite the pessimistic assumptions about the instrument captured in Table~\ref{tbl-instpar} and the large, 16$^\prime$\ coverage of the scan pattern which could be concentrated for smaller targets.  (See discussion at the end of Section~\ref{sec-sim-results}.)}
\end{center}
\end{figure}

Figure~\ref{fig:sens_vs_I} demonstrates how the sensitivity depends on the emission from the galaxy and background. Multiple map pixels can be combined to produce higher signal-to-noise in the measurement of total or polarized intensity, as long as the small noise correlations due to $1/f^n$ noise are taken into account (as is done in Section~\ref{sec-analysis}).  Even though the mapped area extends over most of the 24$^\prime$\ width of the telescope field of view -- larger than needed for a more typical KINGFISH or HRS galaxy -- the sensitivity predicted from the simulation is a factor of 1.6 better than the target outlined in Section~\ref{sec-primapol}.  For a test case with half the linear dimension (and half the velocity and acceleration) of the scan pattern from Figure~\ref{fig:scan_pattern}, we find nearly identical $(I, Q, U)$ maps for the galaxy but an increase of 3.7$\times$ for the integration time per pixel in the center of the map, which leads to a $\sim\! 1.5\times$ improvement in sensitivity in the central 6$^\prime$ diameter.  Using the ``predicted'' instrument parameters rather than the ``modeled'' pessimistic parameters in Table~\ref{tbl-instpar} improves the sensitivity by a further factor of 1.6--1.9.

\subsection{Analysis of Simulated Data \label{sec-analysis}}

\begin{figure*}
\begin{center}
\includegraphics[width=\textwidth]{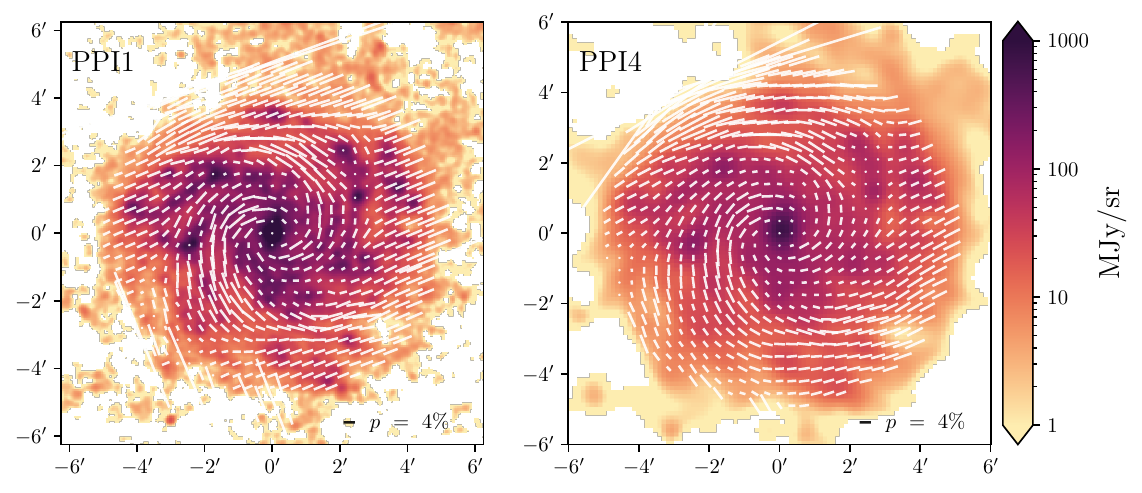}
\caption{The ``observed'' polarization vectors corresponding to magnetic field orientation (i.e., $\psi$ rotated by $90^\circ$) are overlaid on the ``observed'' total intensity maps for PPI1 (left) and PPI4 (right). For visual clarity, one vector in the PPI1 map is an average over a $7\times7$ pixel region (i.e., $21^{\prime\prime}\times21^{\prime\prime}$) while one vector in the PPI4 map corresponds to a $3\times3$ pixel region (i.e., $21^{\prime\prime}\times21^{\prime\prime}$). The vector length is proportional to the polarization fraction.
\label{fig:vector}}
\end{center}
\end{figure*}

\begin{figure*}
\begin{center}
\includegraphics[width=\textwidth]{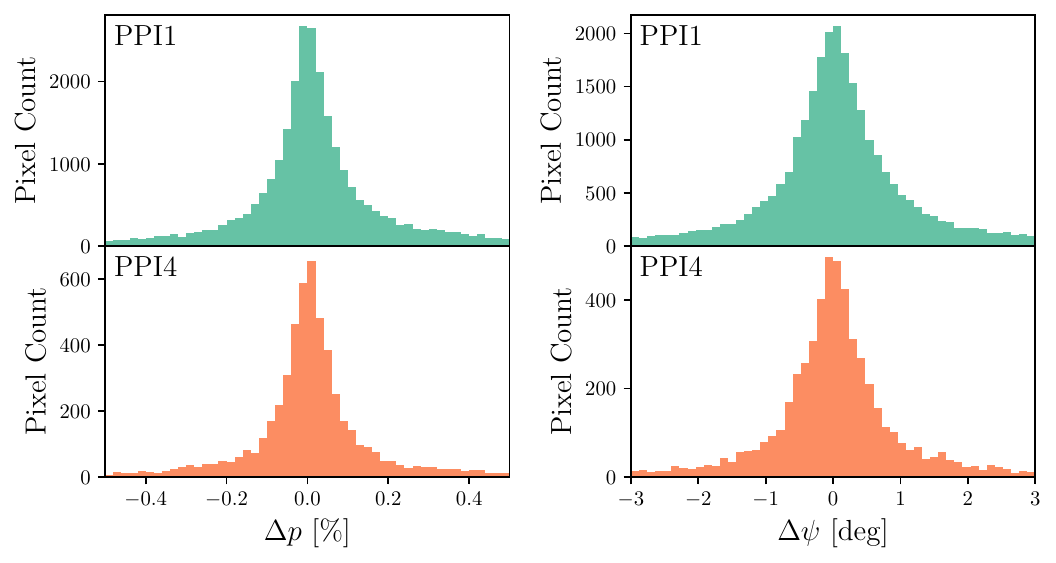}
\caption{The difference in polarization fraction ($\Delta p$) and polarization angle ($\Delta\psi$) between the input signal map and the ``observed'' map in a $10^\prime$ diameter circular region centered on the galaxy, both with zero levels subtracted as described in Section~\ref{sec-analysis}. Both PPI1 and PPI4 are shown at native pixelization, yielding more than five times as many pixels in PPI1.
\label{fig:histograms}}
\end{center}
\end{figure*}

The first step in analyzing the simulated maps is to address the unknown zero level. Since the PPI measurements are purely differential, we follow the simplest approach of finding the constant value for each map that, when added, yields emission consistent with zero in the outskirts of the map. Specifically, we subtract from each of the $I$, $Q$, and $U$ maps the median value of the map within an equal-area annulus surrounding our $10^\prime$-diameter analysis region. We apply the same procedure to the $I$, $Q$, and $U$ ``truth'' maps as well for all comparisons that follow.

The scientific quantities of interest can be derived straightforwardly from the $I$, $Q$, and $U$ maps. First, the polarization fraction $p$ encodes the intrinsic polarization efficiency of the emitting dust grains, the inclination of the magnetic field aligning the grains, and the degree of disorder of the magnetic field along the line of sight. It is given by

\begin{equation}
    p \equiv \frac{\sqrt{Q^2 + U^2}}{I}
    ~~~.
\end{equation}
Because the effects of magnetic field geometry are independent of frequency, the dust composition can be constrained by observing how $p$ varies with frequency. In general, dust models with multiple components predict strong variation of $p$ with frequency especially near the peak of the dust emission spectrum \cite{Draine:2009, Guillet:2018, Draine:2021, Ysard:2024}, whereas models with only one component predict little variation \cite{Hensley:2023}.

The second quantity of interest is the polarization angle $\psi$, which encodes the orientation of the magnetic field aligning the grains as projected onto the plane of the sky. It is given by

\begin{equation}
\psi \equiv \frac{1}{2} \arctan \left(\frac{U}{Q}\right)
~~~.
\end{equation}
Since dust emission is polarized perpendicular to the orientation of the local magnetic field, the measured $\psi$ must be rotated by $90^\circ$ to correspond to magnetic field orientation. Maps of $\psi$ reveal how magnetic fields relate to structures at all scales, while the dispersion in $\psi$ can be used to constrain the local magnetic field strength (e.g., \citenum{planck2016-l11B}). The polarization vectors of the ``observed'' maps are illustrated in Figure~\ref{fig:vector}. Pixels near the edge of the map are comparable in surface brightness to the Galactic cirrus and are thus susceptible to uncertainties in background subtraction. The formal uncertainties on $p$ and $\psi$ are consequently large, but we nevertheless retain them in the plot for illustration.

Figure~\ref{fig:histograms} compares the $p$ and $\psi$ values in the ``observed'' maps to the values in the ``true'' map in a $10^\prime$ diameter circular region centered on the galaxy. While the uncertainty in $p$ and $\psi$ can be straightforwardly computed from the uncertainties in $I$, $Q$, and $U$ under the assumption of Gaussian errors, the presence of $1/f$ noise complicates the picture. We find that 68\% of observed PPI1 pixels and 79\% of observed PPI4 pixels differ by less than 0.2\% in $p$ from the truth maps. For a median $I$ of 38\,MJy/sr and $\sigma(Q) = \sigma(U) = 0.024$\,MJy/sr (see Section~\ref{sec-sim-results}), $\sigma(Q)/I = 0.06\%$, in line with the distribution of values observed here. Likewise, 62\% of observed PPI1 pixels and 73\% of PPI4 observed pixels differ in $\psi$ by less than $1^\circ$. For comparison, SOFIA measurements of nearby galaxies indicate astrophysical dispersion in pitch angles of degrees, if not tens of degrees \cite{Borlaff2023}.

Comparisons of polarization fractions and angles across frequencies in a real analysis would be performed with a common angular resolution, at least as large as the PPI4 beam, which is why the science requirement (Section~\ref{sec-primapol}) is specified per $27^{\prime\prime}\times 27^{\prime\prime}$ resolution element.  The requisite coaddition of pixels enhances the signal-to-noise ratio of the PPI1 maps especially. However, due to pixel-to-pixel covariance, the gains will not be as large as if the pixels were completely independent. We find that the signal-to-noise of the PPI1 map improves by a factor of 6.6 when the pixels are binned to form $27^{\prime\prime}\times 27^{\prime\prime}$ resolution elements, somewhat smaller than the expectation of 9 for uncorrelated noise.  We have applied this analysis to all of the simulations in this paper (using $4\times 4$ binning for PPI4), and the sensitivity results are reported in Table~\ref{tbl-sen_res}.  PRIMAger as simulated meets the polarization science requirements with margins depending on input assumptions.  In the reported cases, the sensitivity refers to an average over the map, and the sensitivity is better than tabulated in the fainter regions.

\begin{table}[h]
\begin{threeparttable}
\caption[]{Summary of polarization sensitivity\tnote{1} results from simulated 1.5 hour observations of NGC 6946}
\label{tbl-sen_res}
\centering
\begin{tabular}{ccccc}
\hline
 & avg. $\sigma(P)$ [MJy/sr], & avg. $\sigma(P)$ [MJy/sr], & avg. $\sigma(P)$ [MJy/sr], & $\sigma(P)$ [MJy/sr], \\
PPI band & pessimistic params. & smaller map\tnote{2} & predicted params. & science requirement\tnote{3} \\
\hline
1 & 0.0067 & 0.0053 & 0.0028 & 0.0069 \\
4 & 0.0067 & 0.0054 & 0.0020 & 0.0069 \\
\hline
\end{tabular}
\begin{tablenotes}
\small
\item [1] The table shows the average uncertainty over the central 10$^\prime$ diameter of the polarized intensity $P$ measured in a $27^{\prime\prime} \times 27^{\prime\prime}$ resolution element.
\item [2] The ``smaller map'' column is measured in a 6$^\prime$ diameter, and the simulation used pessimistic instrument parameters and a smaller scan pattern.
\item [3] The science requirement from Section~\ref{sec-primapol} is scaled from 2.0 hr to 1.5 hr (and from 5$\sigma$ to 1$\sigma$).
\end{tablenotes}
\end{threeparttable}
\end{table}

If in the future PRIMA or a similar mission achieves the sensitivity modeled in this simulation, then it will enable polarization mapping of galaxies $\sim$100$\times$ deeper than SOFIA \cite{Borlaff2023}.  For NGC 6946 specifically, the magnetic field could be mapped, and the dust polarization spectrum measured, throughout the 20 kpc disc for the first time.  Smoothed to beam scale, the simulated PPI4 polarized intensity maps reach $\sim$5 kJy/sr or better, within a factor of $\sim$2 of the extragalactic polarized intensity background fluctuations \cite{Bethermin2024}.

\section{Relative Gain Calibration\label{sec-gain}}

Good calibration of the relative detector gains $\{g_i/g_j\}$ is necessary in a case such as PPI in which multiple detectors will be used, without polarization modulation, to recover the polarization signals.  First, we demonstrate that the relative gains can be derived from observations of an astronomical source, even if it has intrinsic polarization, as long as it is observed for three line-of-sight rotation angles separated by tens of degrees.  Absolute flux density calibration is not discussed here, so the total intensity $I$ is divided out:

\begin{align}
\label{eq-sigmodel2}
S_{i,j,k} &= g_i I_{p(i,j,k)}[1 + \epsilon_i q_{p(i,j,k)} \cos 2(\theta_i-\phi_j) + \nonumber \\ &\epsilon_i u_{p(i,j,k)} \sin 2(\theta_i-\phi_j)]
\end{align}
The baseline $b$ is dropped here, presumed to have been determined and subtracted as in Section~\ref{sec-destripe}, and the observation is presumed long enough to make the noise $n$ negligible.  $q$ and $u$ give the fractional polarization expressed in Stokes parameters.

We consider a highly simplified observational example, consisting of two steps.  First, a single source position is observed with a single detector at three rotation angles.  This provides three measurements, with three unknowns:  the average response $g_1 I_{p1}$, $q_{p1}$, and $u_{p1}$.  As long as there is sufficient range in $\phi_j$ (ideally $\sim$90$^\circ$ for maximum signal-to-noise), the set of three instances of Equation~\ref{eq-sigmodel2} can be solved for the unknowns.  Next, an additional observation of the same source position is made with a second detector.  With all of the terms inside the square brackets now known, the average response $g_2 I_{p1}$ is derived, and therefore $g_1 / g_2$.  This process could be extended to the full detector array; in practice, the relative gains will be derived iteratively with the source map and baselines over a large data set with significant line-of-sight rotation and redundant observations of a field.

Line-of-sight rotation for PRIMA is achieved by re-observing a target some number of days, weeks, or months later.  The PRIMA Sun/Earth/Moon avoidance constraints permit only modest  rotation for sources at low ecliptic latitude.  However, targets within 10$^\circ$ of the ecliptic poles are continuously viewable and can be observed with full rotation of the field.  The Large Magellanic Cloud is 5$^\circ$ from the south ecliptic pole and is bright and extended in the far infrared \cite{Meixner:2013}, making it an excellent candidate for PPI relative gain calibration.  To ensure that the detector system gain is stable over multiple, rotated observations, measurements are made of the internal calibration source, designed to produce a temporally-constant illumination over similarly long periods.

An error in relative gain for a detector will leave an imprint (positive or negative) in the $Q$ and/or $U$ maps, proportional to Stokes $I$.  For a reference polarization systematic error limit of 0.5\%, the relative gains need to be known to better than $\sim$0.7\% accuracy for measurement with a minimum number of detectors.  However, this systematic effect averages down as more detectors with uncorrelated gain errors measure a given point on the sky.  Relative gain errors can also lead to errors in the baseline determination and therefore low-level but undesirable stripes in the maps radiating from brighter sources.  These are reduced by setting an upper intensity limit for samples which are used in the baseline fits.

To assess the calibration accuracy requirement for PPI, we performed a PPI1 observation simulation as described in Section~\ref{sec-sim}, but this time, relative gain error is introduced and implemented as a gaussian random distribution over the detectors with mean of 1 and standard deviation of 0.05.  We evaluated results by performing comparison of polarization fraction p and angle as in Figure~\ref{fig:histograms}, with the difference now being for the simulation with gain error vs. one with no gain error.  We find that even with the 5\% (1$\sigma$) gain error as described, the 1$\sigma$ error in $p$ is still $<$0.5\% .

\section{Pointing and Beam Matching Requirements\label{sec-beam}}

In this last section, we outline systematic effects related to matching of detector beams (spatially or temporally) and how they are addressed in PPI observations.  In all cases, the primary concern is conversion of total intensity $I$ into an erroneous, artifact $Q$ or $U$ with a spatial dependence.

First, we consider the relative pointing knowledge requirement.  In the single-polarization detector approach such as in PPI, the detector differencing to measure polarization is not done with simultaneous measurements, so a drift in pointing between the measurements can lead to a residual in the polarization.  In the limit of bounded pointing error and many redundant measurements, the residual will tend to average down with time, with degraded image resolution becoming the lasting effect.  Below, we consider a worst case applicable to sparse sampling, to place a lower limit on the relative pointing knowledge.

\begin{figure*}
\begin{center}
\includegraphics[width=\textwidth]{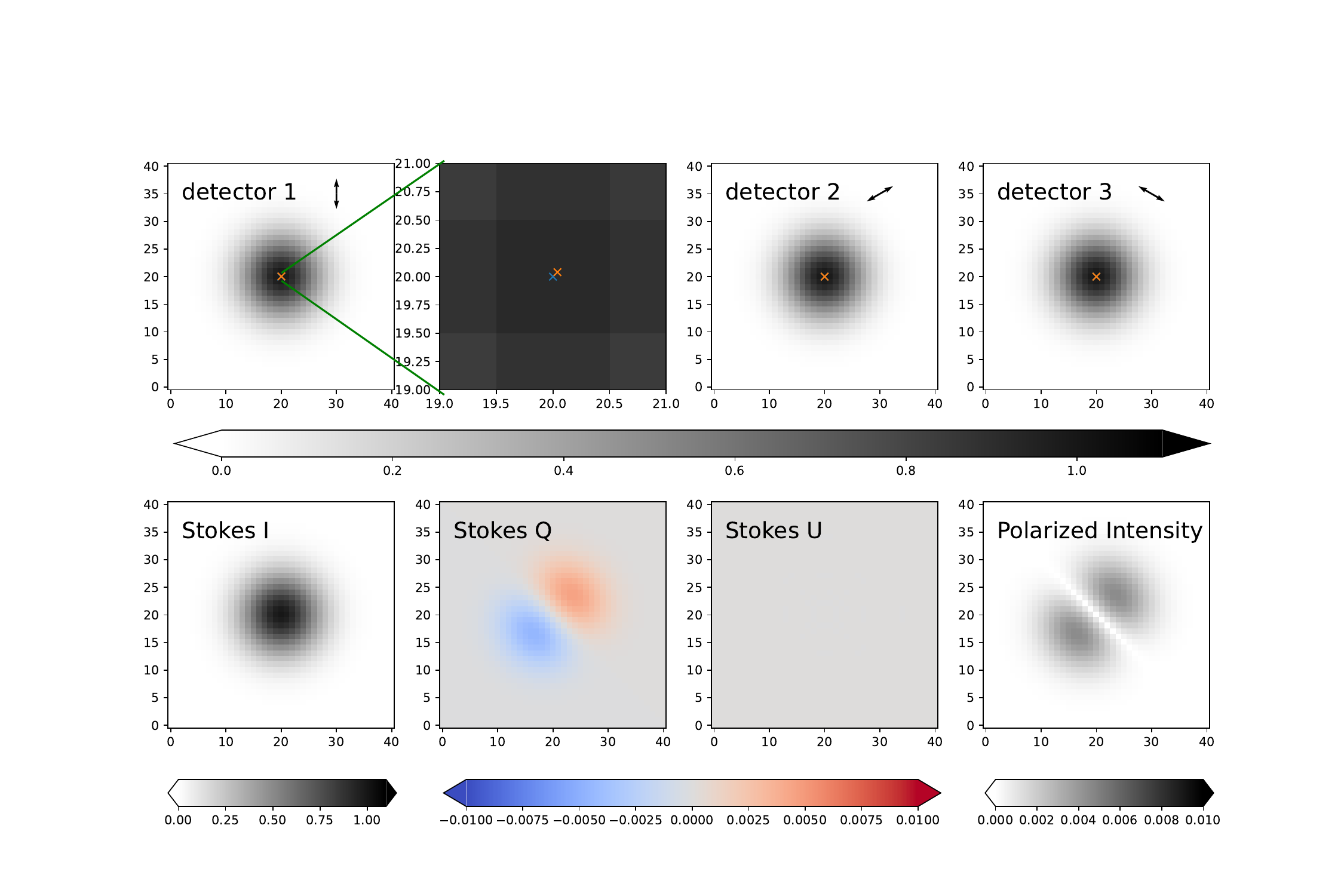}
\caption{Demonstration of a polarization residual caused by relative pointing error. The top row shows the response of three detectors to an unpolarized, unresolved source.  The three detectors have polarization response directions as shown by the vectors, and they are combined to solve for $(I,Q,U)$.  The source as observed by detector 1 is displaced by 0.005 beam FWHM (as indicated by the red and blue X's in the inset) to simulate relative pointing knowledge error for that detector.  The bottom row shows the resulting polarization artifacts, with a peak in polarized intensity of 0.5\%$\times$ the peak in total intensity.\label{fig:point_err}}
\end{center}
\end{figure*}

For observation of an unpolarized, unresolved source, pointing displacement by $\sim$0.005 beam FWHM causes a 0.5\% residual in the polarized intensity map (Figure~\ref{fig:point_err}), which sets a relative pointing knowledge target for the observations.  At the wavelength of PPI1, this is 0.06$^{\prime\prime}$ . 

Error in the pointing information provided in the PRIMA telemetry can be lessened by using the PPI measurements themselves to make corrections.  This technique has been demonstrated qualitatively using, e.g., Herschel/SPIRE 250--500 $\mu$m imaging\cite{Stone2016}.  It is difficult to forecast the kind of pointing error that PPI will experience; potentially, $\sim$4 additional fit parameters could be added per raster line, to correct starting position and velocity.

Intensity gradients in the observed field enable relative alignment of the detectors sampling it.  Order-of-magnitude, the uncertainty in the measured alignment shift $(\Delta x, \Delta y)$ of $N$ detector samples is estimated by:
\begin{equation}
\sigma(\Delta x) \approx \frac{\sigma(I)}{\sqrt{N}\sqrt{<|\frac{dI}{dx}|^2>}} ,
\sigma(\Delta y) \approx \frac{\sigma(I)}{\sqrt{N}\sqrt{<|\frac{dI}{dy}|^2>}}
\end{equation}
where $\sigma(I)$ is the uncertainty in a single intensity measurement and $\sqrt{<|\frac{dI}{dx}|^2>}$ is the root-mean-square intensity gradient in the $x$ direction.  For the central 6$^\prime$\ diameter of NGC 6946, in the PPI4 band, the rms intensity gradient in one direction is $\sim$2.5 MJy/sr/arcsec .  For the mean intensity of $\sim$90 MJy/sr in this region, the measurement uncertainty is $\sim$1.9 MJy/sr per detector per 350 Hz sample.  If $N \approx 30$ detectors overlapping the galaxy are used to measure the pointing, then the uncertainty in alignment (in 2D space) is $\sim$0.2$^{\prime\prime}$, or 0.007$\times$ the beam FWHM.  This is near the pointing alignment target from earlier in this section, and the combination of multiple samples further improves the measurement.

A second effect is due to beam shape variation.  Practicalities of optical design and instrument assembly mean there will be some variation in beam width and ellipticity across the field of view for a given PPI band.  Another possible type of beam shape variation is beam ellipticity which aligns with the detector polarization direction.  In either case, the detector differencing to derive polarization will show spatially-dependent residuals in $Q$ or $U$ resulting from differential response to $I$.  This effect does not average down unless the scan pattern is changed in such a way that the patterns of residuals are changed in orientation or amplitude.

Differential width in the beam among the detectors measuring $(I,Q,U)$ creates an annular polarization residual around an unpolarized unresolved source, and $\sim$1\% fractional difference in width causes a 0.5\% peak in the polarized intensity.  Differential ellipticity with direction uncorrelated with detector polarization orientation produces a quadrupolar residual in polarized intensity with similar sensitivity:  a $\sim$1\% fractional difference causes a 0.5\% peak in the polarized intensity.  Ellipticity aligning with the detector polarization produces more of an annular pattern.

Two mitigations for the beam shape effects are envisioned for PPI.  One is implemented by the focal plane design, which has the three polarization angles well mixed and therefore guards against large-scale variations in beam shape across the focal plane.  The other mitigation is implemented in the science data pipeline, if needed.  A pipeline step smooths the measured data, using a convolution kernel customized to each detector, in order to better match the beam shapes in the maps.  This smoothing occurs before the last solution for the $(I, Q, U)$ maps.  It remains to be seen, from better knowledge of beam shapes and further simulation, whether the convolution step can be postponed until after the signal baselines (Section~\ref{sec-destripe}) are derived, or the convolution will need to be performed for each iteration.

A third potential beam effect is a different shape of the beam in the cross-polar direction.  Ideally, there is no response for light polarized perpendicularly to the co-polar direction of the detector.  However, for polarization efficiency $\eta < 1$, there is a weak response with likely a significantly different beam shape.  For an unpolarized source, the cross-polar response produces only a slight modification of the beam shape — an asymmetry oriented with respect to the detector co- and cross-polar directions.  The residuals in polarization are as described for differential beam width and ellipticity above.  For a polarized source, the cross-polar response causes a polarization-dependent beam shape.  This effect is expected to be negligible for achieving PRIMA's primary polarimetry science, but it will be studied in more detail in the future if needed.

\section{Conclusions\label{sec-conclusion}}

In this paper, we have explored the performance of the polarimetric imager that is proposed for the PRobe far-Infrared Mission for Astrophysics (PRIMA). Its key features are that each focal plane is filled with single-polarization detectors equally sampling three linear polarization directions (120$^{\circ}$ apart), and that it is not using a half-wave plate to modulate the incoming polarization. Using a realistic input sky, based on the driving polarization science case for PRIMA, and an observing strategy that makes use of the agile internal beam steering mirror, we demonstrate that even with pessimistic assumptions about the noise properties of the detectors, including its $1/f$ component, a least-squares reconstruction approach recovers the source polarized and total intensity to an accuracy that satisfies the mission requirements.

More precisely, we find that reconstructed maps suffer only a 30-40\% increase in noise level compared to the case of pure gaussian noise. For a simulated 1.5 hour observation of the nearby galaxy NGC 6946, the polarized emission is detected with high signal-to-noise, both in the galaxy and cirrus foreground. In the region of interest in the simulation, the fidelity of the reconstructed information is excellent.
Our simulations also show that PPI out-performs the sensitivity requirements set by the mission by a factor of 1.6--5 (depending on assumptions for the instrument and scan pattern), which are orders of magnitude better than what has been demonstrated to date.

The design choice of single-polarization detectors with no polarization modulator places requirements on calibration of certain aspects of the instrument and observation.  We considered the effects of detector gain error, pointing error, and beam shape mismatch, outlining strategies for their measurement and placing limits on the magnitude of the effects following correction.

While this will need to be demonstrated by relevant simulations, we expect to draw similar conclusions for a General Observer science case of PRIMA polarimetry, that of measuring magnetic fields and the nature of dust in degree-sized star-forming clouds \cite{Moullet:2023}. Here as well, high-frequency modulation by the BSM, superimposed on slower spacecraft scanning motion, should allow the approach developed here to result in accurate and efficient map reconstruction.

\section*{Disclosures}
The authors declare that there are no financial interests, commercial affiliations, or other potential conflicts of interest that could have influenced the objectivity of this research or the writing of this paper.

\subsection*{Code, Data, and Materials Availability} 

The code (Python notebook) used in this research will be made available from the corresponding author by email request.

\section*{Acknowledgments}
We acknowledge valuable feedback on the paper draft, simulator, and measurement approach provided by Jochem Baselmans, Matt Bradford, David Chuss, Lorenza Ferrari, Laura Fissel, Jason Glenn, Willem Jellema, Margaret Meixner, and Stephen Yates.

This research was carried out in part at the Jet Propulsion Laboratory, California Institute of Technology, under a contract with the National Aeronautics and Space Administration.


\bibliography{refs, refsPlanck}
\bibliographystyle{spiejour}   

\vspace{2ex}\noindent\textbf{C. Darren Dowell} is a scientist at the Jet Propulsion Laboratory, California Institute of Technology.  He received a BA in physics (space physics and astronomy) from Rice University in 1991 and MS and PhD in astronomy and astrophysics from University of Chicago in 1992 and 1997, respectively.  He has led several instrument projects for imaging and polarimetry at infrared/(sub)millimeter wavelengths, most recently the HAWC+ imaging polarimeter for SOFIA.  His current research interests include infrared astrophysics and data analysis methods.

\vspace{2ex}\noindent\textbf{Brandon S. Hensley} is a scientist at the Jet Propulsion Laboratory, California Institute of Technology. He received his BS in physics from Caltech in 2010 and his MA and PhD degrees in astrophysical sciences from Princeton in 2012 and 2015, respectively. His current research interests include interstellar dust, polarimetry, and the cosmic microwave background.

\vspace{2ex}\noindent\textbf{Marc Sauvage} is an astrophysicist at the Commissariat \`a l'\'Energie Atomique et aux \'Energies Alternatives (CEA-Saclay). He has been an instrument scientist on Herschel/PACS and the Ground Segment Scientist on Euclid. His research interests revolve around star formation and interstellar medium processes.

\listoffigures
\listoftables

\end{spacing}
\end{document}